\documentclass{aa} 

\usepackage{graphicx}
\usepackage{natbib} 
\usepackage{txfonts} 
\begin{document}

   \title{Satellite observation of the dust trail of a major bolide event over the Bering Sea on December 18, 2018}

   \author{J. Borovi\v{c}ka 
          \inst{1}
          \and
          M. Setv\'ak\inst{2} 
                   \and
         H. Roesli\inst{3} 
          \and
         J. K. Kerkmann\inst{4}
          }

   \institute{Astronomical Institute of the Czech Academy of Sciences, Fri\v{c}ova 298, CZ-25165 Ond\v{r}ejov, Czech Republic \\
              \email{jiri.borovicka@asu.cas.cz}
         \and
             Czech Hydrometeorological Institute, Satellite Department, Na \v{S}abatce 17, CZ-14306 Praha 4, Czech Republic\\
             \email{martin.setvak@chmi.cz}   
                  \and
          Locarno, Switzerland,   \email{satmet.hp@ticino.com}  
        \and
          European Organisation for the Exploitation of Meteorological Satellites, EUMETSAT Allee 1, D-64295 Darmstadt, Germany       }

   \date{Received 11 September 2020; accepted 19 October 2020}

 
  \abstract
{One of the most energetic bolide events in recent decades was detected by the US Government sensors (USGS) 
over remote areas of the Bering Sea on December 18, 2018, 23:48 UT.
No ground-based optical observations exist.}
{Using the satellite imagery of the dust trail left behind by the bolide, we tried to reconstruct the bolide trajectory. In combination with the bolide speed
reported by the USGS, we computed the pre-atmospheric orbit. 
Observations in various spectral bands from 0.4~$\mu$m to 13.3~$\mu$m enabled us to study the dust properties. }
{Images of the dust trail and its shadow obtained from various angles by the Multi-angle Imaging SpectroRadiometer (MISR) on board the Terra polar satellite and
geostationary satellites Himawari-8 and Geostationary Operational Environmental Satellite 17 (GOES-17)
were used. The initial position and orientation of the trail was varied, and its projections
into the geoid coordinate grid were computed and compared with real data. 
Trail motion due to atmospheric wind was taken into account. Radiances and reflectances
of selected parts of the dust trail were taken from the Moderate-resolution Imaging Spectroradiometer (MODIS) on board Terra. 
Reflectance spectra were compared with asteroid spectra. }
{The bolide radiant was found to be $13\degr \pm\ 9 \degr$ from that reported by the USGS, at azimuth 130\degr\ (from south
to west) and zenith distance 14\degr. The bolide position was confirmed, including the height of maximum dust deposition around 25 km.
The incoming asteroid had to be quite strong to maintain a high speed down to this height. The speed of 32 km s$^{-1}$, reported by the USGS,
was found to be plausible. The orbit had a high inclination of about 50\degr\ and a perihelion distance between 0.95--1 AU. The semimajor axis
could not be restricted well but was most probably between 1--3 AU. The dust reflectance was much lower in the blue than in the red,
consistent with the material of A- or L-type asteroid. The absorption at 11~$\mu$m confirms the presence of crystalline silicates in the dust.}
{}


   \maketitle
%

\section{Introduction}

Impacts of cosmic bodies, asteroids and comets, played an important role in the history 
of life on the Earth. Fortunately, impacts of bodies large enough to have severe global
consequences occur only on geological timescales. In historical times, the largest recorded impact  
was the Tunguska event of June 30, 1908 \citep[see, e.g.,\ ][and references therein]{Tunguska}. 
Its energy was estimated 
to be about 10 Mt TNT (1 kt TNT = $4.185 \times 10^{12}$ J) and it was caused by 
an impactor of a diameter 40 -- 80 meters \citep{Tungus-energy}. More recently, 
on February 15, 2013, a 19-meter asteroid with an energy of 500 kt TNT exploded near the
city of Chelyabinsk \citep{Chelya_Brown, Chelya_Popova}. This event was remarkable
because it occurred in a densely populated region. Significant damage was caused by the associated
blast wave, and about 1600 people were injured, mostly lightly \citep{injuries}. 
The Chelyabinsk bolide was well documented by hundreds of casual videos \citep{videos},
as well as by seismic and infrasonic detectors \citep{Chelya_Popova, Chelya_Brown}.
The massive dust trail left in the atmosphere was also visible from satellites in the Earth's orbit
\citep{Miller, Proud}.

According to current estimates, Tunguska-like impacts occur on average once per 500 years
and events like in Chelyabinsk once per 50 years \citep{Harris, Tricarico}. 
Bolides with energy of 100 kt TNT, caused by $\sim$10-meter asteroids, appear once in a decade on average.
Asteroids of sizes 1--20 m belong to the least known bodies of the inner Solar System.
Their impacts produce very bright bolides \citep[so-called superbolides,][]{superbolides}, which typically only last several seconds, however.
Nearly continuous monitoring of 
the atmosphere is therefore needed to detect them.
No scientific system is devoted to monitoring superbolides on a global scale.
They are observed as a byproduct by the US Government sensors (USGS), which is 
a military system \citep{Tagliaferri, BrownUSGS}.
The detections are currently being published online in a table\footnote{https://cneos.jpl.nasa.gov/fireballs/}.
The published data are limited to the date and time of peak brightness, latitude and longitude with a
precision of 0.1\degr, velocity components, total radiated energy, and total
impact energy calculated from the radiated energy by the empirical formula of \citet{Brown2002}.
Not all detected bolides are included, and for some bolides, the information is incomplete.
Moreover, the data, especially velocities, were found to be unreliable when verified by
independent ground-based observations \citep{Romanian, Devillepoix}. 

Recently, the Geostationary Lightning Mappers (GLM) on board two Geostationary Operational Environmental 
Satellites (GOES-16 and 17)  have proved to be capable
of detecting bolides as well \citep{GLM}, but their coverage is limited to the Americas
and the adjacent oceans. GLMs provide detailed light curves in a narrow passband, making the estimate of the total
radiated energy difficult. Trajectories and velocities are not provided.

Another global system is the network of infrasound stations from the International Monitoring System (IMS) 
operated by the Comprehensive Nuclear-Test-Ban Treaty Organization (CTBTO),
which also detects bolides as a byproduct \citep{CTBTO}. The data are not routinely scanned
for bolides but are rather used to check bolide reports based on other means. Infrasonic data can
provide a rough position of the bolide and an independent energy estimate based on an empirical formula
relating the period of the signal with the source energy \citep{infrasound}.

Except for Chelyabinsk, the most energetic bolide event of the last several decades reported
until recently was the Indonesia bolide of October 8, 2009, with an estimated energy of about 50 kt TNT \citep{Indonesia}. 
The USGS table lists 33 kt TNT for this event, and similar energies are also listed for the 
Marshall Island bolide of February 1, 1994 \citep[see also][]{Marshall}, and the December 25, 2010, event
east of Japan \citep[see also][]{Dec25}. 

It was a surprise when in March 2019, a report of a 173 kt TNT bolide appeared on the 
USGS page. The reported event had occurred already on December 18, 2018, over the remote Bering Sea, and 
remained otherwise  unnoticed. Its signal was subsequently found on 16 infrasonic stations (P. Brown, private comm.).

Because of the rarity of such an energetic event, it was desirable to obtain as much information
as possible. 
The bolide occurred over sea, 350--500 km from the nearest
and sparsely populated coast of the Russian peninsula Kamchatka and from some islands in the Bering Sea, under cloudy weather, 
and during daytime. Evidently, nobody saw the bolide or recorded it
on a camera from the ground.  No reports appeared from airplane pilots passing by the area either. 
Thus, the only chance for additional information are satellite observations.

The location of the bolide occurrence was outside the area covered by the GLM detectors. Other Earth-observing
instruments had a low chance to detect the bolide itself. Nevertheless, the Japanese
geostationary satellite Himawari imaged the bolide dust trail and its temporal evolution. Under
somewhat poor conditions, the dust trail was also imaged by  GOES-17. 
Fortunately, the polar orbiting satellite Terra overpassed the region just five
minutes after the bolide and documented the dust trail from a closer vicinity. Other polar
satellites, the National Oceanic and Atmospheric Administration 20 (NOAA-20),
and the Suomi National Polar-orbiting Partnership (Suomi-NPP), detected signs of the dispersed dust trail one hour 
and almost two hours later, respectively.

This paper is devoted to the analysis of the dust trail images, primarily from the Terra and Himawari satellites.
The aim is an independent determination of the bolide trajectory and the description of the dust trail. 
Bolide dust clouds have been observed from the orbit before \citep{Klekociuk, Charvat}, but
only Chelyabinsk showed a well-defined dust trail suitable for trajectory determination 
\citep{Proud}. While numerous ground-based records
in Chelyabinsk enabled more precise trajectory (and velocity) determination \citep{Chelya_Borovicka,
Chelya_Popova}, the satellite data are the only option in case of the Bering Sea bolide.

The available data are described in Section~\ref{data}.
In Section~\ref{trajectory} the determination of the trajectory from dust trail images is explained in detail.
The most likely orbit of the bolide is also discussed.
The trail evolution as observed from Himawari is briefly described in Section~\ref{evolution}. 
Section~\ref{spectrum} is devoted to the analysis of the dust from observations in different spectral bands.
All results are summarized and discussed in Section~\ref{discussion}.

   \begin{figure*}
   \centering
   \includegraphics[width=18cm]{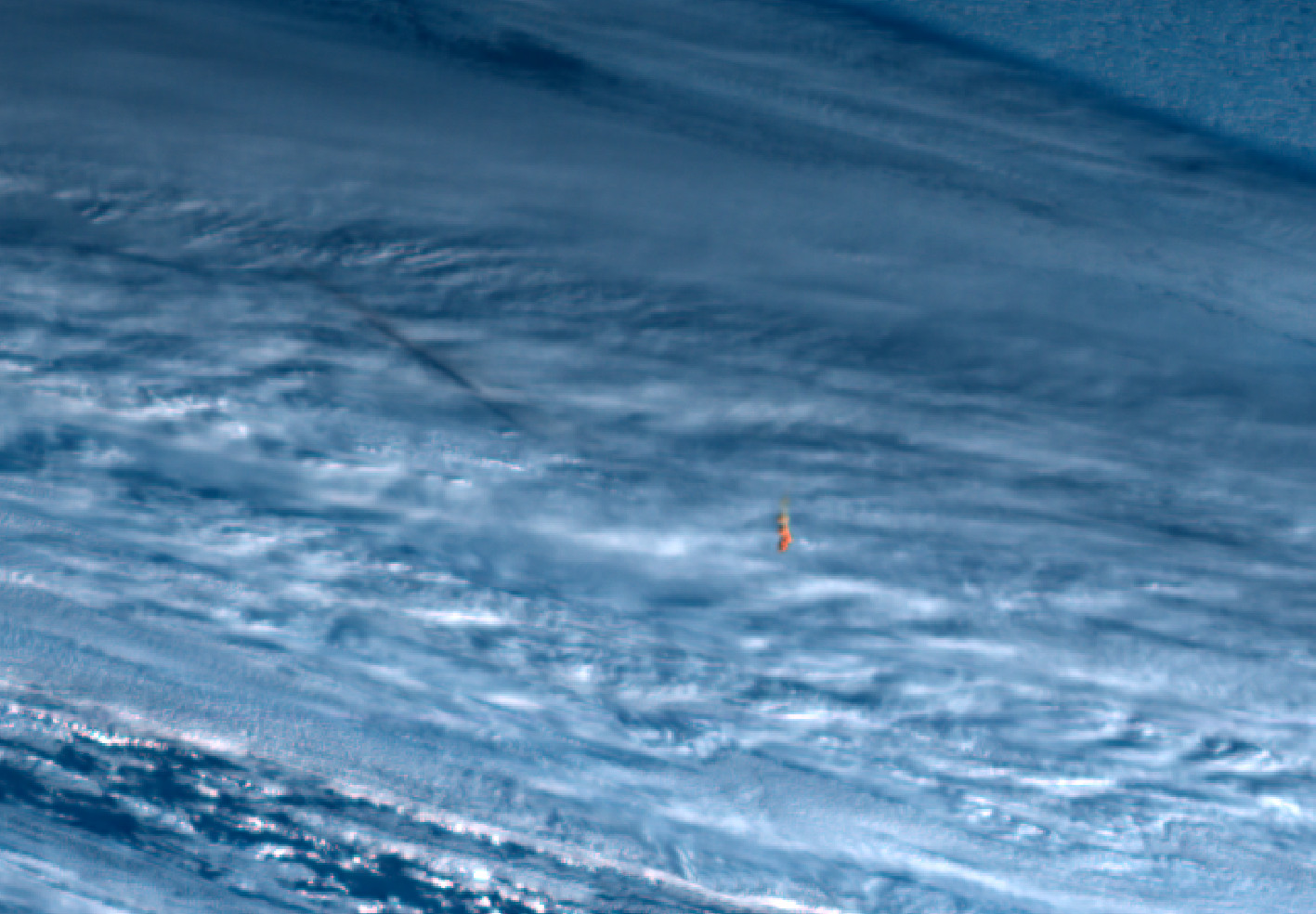}
   \caption{True-color red-green-blue (RGB) composite image of the bolide dust trail (orange object in the middle) and its shadow cast on the cloud cover
   (upper left) taken by the geostationary Himawari-8 satellite at nominal time 23:50 UT. 
   Section from a full-disk image in native projection.}
              \label{Himawari}%
    \end{figure*}

   \begin{figure}
   \centering
   \includegraphics[width=8.5cm]{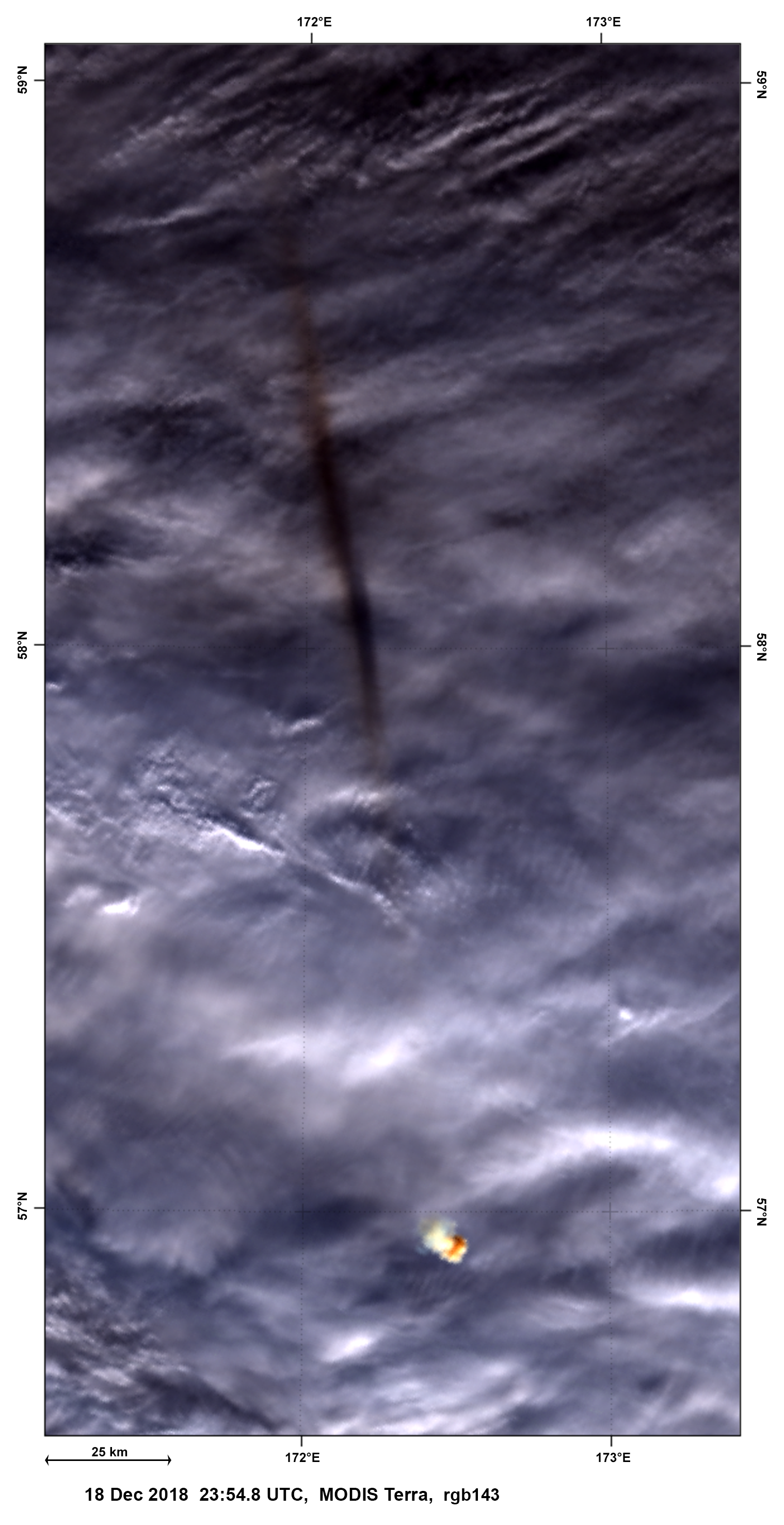}
   \caption{RGB composite image of the bolide dust trail and its shadow taken by the MODIS 
   instrument on board the polar orbiting satellite Terra at 23:55 UT. 
   The image was rotated and remapped to transverse Mercator projection. North is up. The geographical
   coordinates (valid for zero altitude) are indicated at the side.}
              \label{MODIS}%
    \end{figure}
    
       \begin{figure}
   \centering
   \includegraphics[width=7.5cm]{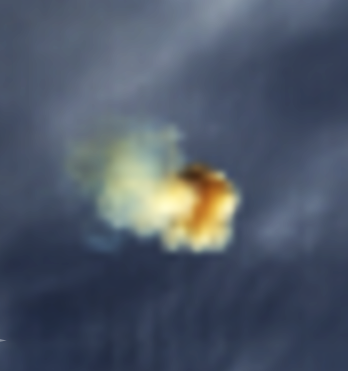}
   \caption{Detail of the trail image from Fig.~\ref{MODIS}. To avoid remapping artifacts, this image was left in the original 
   satellite swath projection. Bands 1 (646 nm), 4 (554 nm), and 3 (466 nm) were used for the RGB colors.
   Color balance was made in such a way that the brightest regular clouds nearby appear white. 
   The zoom factor was 4$\times$ for band 1 (250 m resolution) and 8$\times$ for the other two bands (500 m resolution).
   Resampling was made using the optimized bicubic (OBC) method of the ENVI 
   software (https://www.harrisgeospatial.com/Software-Technology/ENVI).}
              \label{detail}%
    \end{figure}

\section{Available data}
\label{data}

\subsection{US Government sensors}
\label{USGS}

The USGS data (retrieved in April 2020) give the data and time of the bolide as
2018-12-18, 23:48:20    UT and the location of the peak brightness as 56.9\degr\ northern
latitude, 172.4\degr\ eastern longitude, and a height of 25.6 km. The velocity
at peak brightness was listed as 32.0 km s$^{-1}$ with the components in the  
Earth-centered-Earth-fixed (ECEF) system $v_x=6.3$ km s$^{-1}$, $v_y=-3.0$ km s$^{-1}$, 
and $v_z=-31.2$ km s$^{-1}$. It corresponds to the azimuth of the radiant (counted
clockwise from south to west) of 169.4\degr\ and zenith distance 21.4\degr. 
According to these data, the bolide moved on a steep trajectory nearly
from north to south. The corresponding equatorial coordinates of the apparent radiant are
Right Ascension = 239.2\degr, Declination = 77.4\degr. The reported total radiated energy
was $1.3\times10^{14}$ J and the total impact energy was calculated to be 173 kt TNT.

The velocity of 32 km s$^{-1}$ is unusually high. Of the 59 meter-scale impactors studied
by \citet{BrownUSGS}, mostly on the basis of USGS data,  only 2 had a velocity of
32 km s$^{-1}$ or higher. Moreover, the fastest bolide in that sample, the January 7, 2015,
bolide over Romania, was observed from the ground and the true velocity was found to be 
$27.79 \pm 0.19$ km s$^{-1}$, instead of 35.7 km s$^{-1}$ reported by the USGS
\citep{Romanian}. The radiant was off by seven degrees. In other cases, for example, Buzzard Coulee,
the USGS underestimated the velocity \citep{BrownUSGS}.

If the pre-impact heliocentric orbit of the Bering Sea bolide  is computed using the USGS data, 
a clearly cometary-type orbit is obtained with semimajor axis $a=4.0$ AU, aphelion distance $Q=7.0$ AU, 
and inclination $i=48\degr$. Using the simplest criterion for an orbit to be asteroidal,
$Q<4.6$ AU \citep{Kresak}, we found that it would be sufficient to lower the velocity to 30.2 km s$^{-1}$
to obtain an asteroidal orbit. However, it is also possible that the radiant is incorrect. 
Alternatively, the object could indeed be cometary. To obtain more clues, we analyzed
the dust trail images.

   \begin{figure*}
   \centering
   \includegraphics[width=13.5cm]{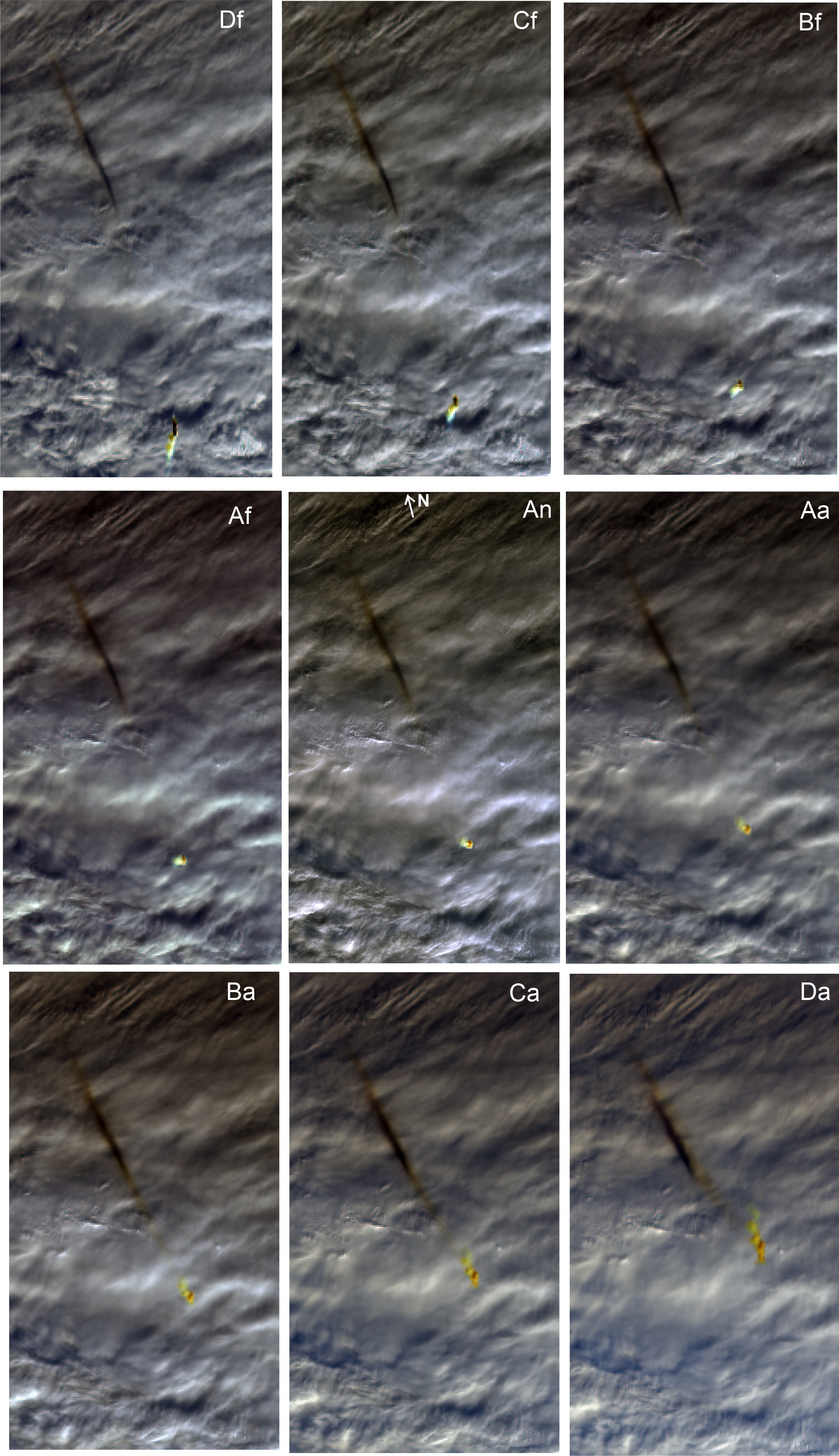}
   \caption{Nine RGB composite images of the bolide dust trail and its shadow taken by nine
   different cameras of the MISR instrument on board the polar orbiting satellite Terra. 
   The first image, acquired by the forward-looking camera Df, is in the upper left corner. The sequence continues
   along the rows until the last image, acquired by the backward-looking camera Da in the bottom right corner.
   The images were remapped to the WGS84 ellipsoid using the space oblique Mercator projection. 
   The vertical axis follows the satellite path; north is indicated in the An image.}
              \label{MISR}%
    \end{figure*}

   \begin{figure}
   \centering
   \includegraphics[width=8.5cm]{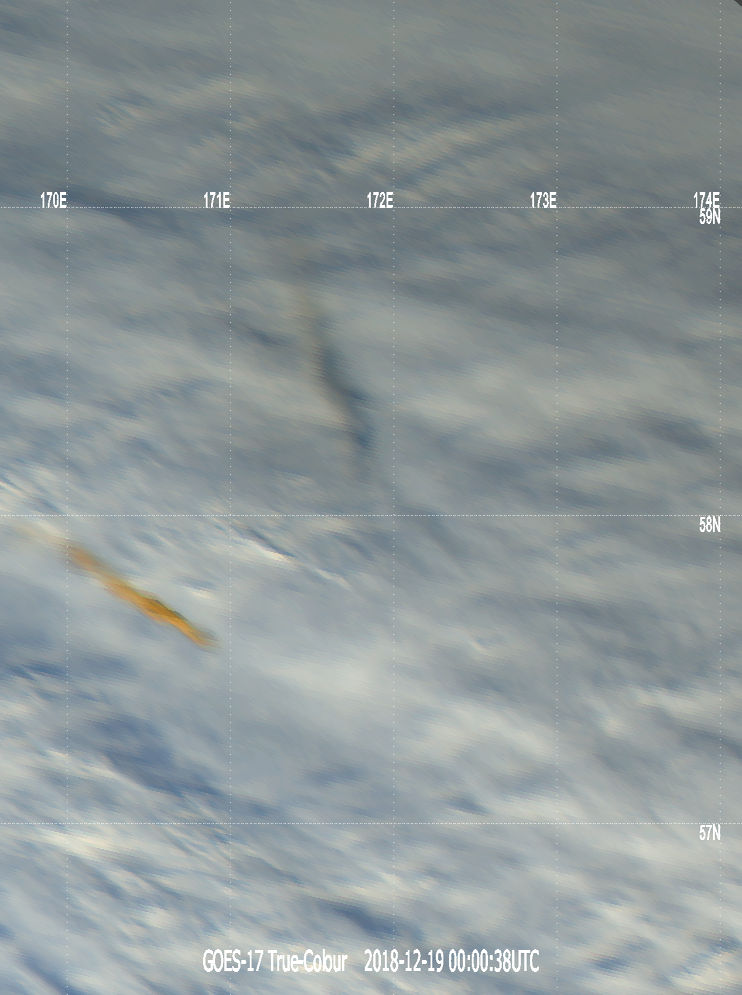}
   \caption{True-color RGB composite enhanced for good contrast of the bolide dust trail (left) and its shadow 
   taken by the geostationary GOES-17
   at nominal time 00:00 UT. The view was remapped from full disk into local latitude-longitude grid. The grid is valid for 
   zero altitude. Both trail and shadow are parallax shifted relative to each other, and also with respect to sea level.}
              \label{GOES}%
    \end{figure}

\subsection{Satellite images of the fresh dust trail}

The bolide occurred at high northern latitude in mid-northern winter, but shortly before
local noon. The scene was therefore illuminated by the Sun located low above the southern horizon
(the center of the Sun was 9\degr\ above horizon at azimuth $-9\degr$ for a ground-based
observer at the location of the bolide)\footnote{Computed with the NASA JPL HORIZONS system, 
https://ssd.jpl.nasa.gov/?horizons.}.

Figure~\ref{Himawari} shows the view from the Himawari-8 satellite. Himawari-8 is a Japanese weather satellite
operated by the Japan Meteorological Agency. It is located on geostationary orbit above 140.7\degr\ eastern
longitude. The image was taken by the Advanced Himawari Imager (AHI). 
The dust trail first appeared in the image with a time stamp 23:50 UT. AHI works in scanning mode and
scans the full disk every 10 minutes in 16 spectral bands, starting from north. 
The scanning pattern is rather complicated because 
scanning of the full disk is combined with more frequent scanning of areas of interest\footnote{See 
https://www.data.jma.go.jp/mscweb/en/himawari89/space\_segment/ spsg\_ahi.html}. 
It follows that the region of the dust trail was scanned about 35 s after the scan beginning, that is,\ at about
23:50:35 UT, nearly two minutes after the bolide.

The image shows a reddish dust cloud elongated in vertical direction hovering above regular clouds. 
A long narrow shadow can be seen on the clouds toward the north. The length of the shadow indicates
that the dust had a significant vertical extent and can indeed be called a dust trail. The large gap between the
dust trail and its shadow indicates that the trail was much higher in the atmosphere than the top of the clouds.

Figure~\ref{MODIS} shows the view from the Terra satellite. Terra is a National Aeronautics and Space Administration (NASA) Earth observing satellite
on circular Sun-synchronous polar orbit with inclination of 98.5\degr, period 99 minutes, and altitude 705 km
above surface. The image was taken by the Moderate-resolution Imaging Spectroradiometer (MODIS). 
As the satellite circles the Earth, MODIS scans a 2330 km wide swath in 36 spectral 
bands\footnote{See https://modis.gsfc.nasa.gov}. The resolution is from 250 meters to 1 km at nadir, depending
on the spectral band.  At the time of the bolide, the satellite was located over the Arctic Ocean and was heading nearly south
toward the Chukotka peninsula and eastern Australia. Nearly six minutes later, at 23:54:51 UT, 
the satellite passed at only 50 km horizontal distance to the west of the bolide location. 
The MODIS view therefore represents an almost nadir view of the dust trail. Because the trail was nearly vertical,
it was greatly foreshortened in this perspective and looks like a single blob. The shadow, on the other hand, 
is about 100 km long. Most of the trail appears bright yellow, while the shadow is dark brown.

Figure~\ref{detail} is a zoomed view of the dust trail. Two parts of the trail can be distinguished. 
The part to the right (southeast) shows yellow and brown dust. In the lower part of the 
trail, the dust concentration was higher. The yellow regions were directly illuminated by the Sun, while
the brown regions were partly shadowed. The part to the left (northwest) appears brighter with yellow-bluish color. 
In the upper part of the trail with lower dust concentration,  the sunlight was scattered.

Another instrument on board the Terra satellite is the Multi-angle Imaging SpectroRadiometer (MISR).
Unlike MODIS, MISR does not scan only across nadir, but also has four forward-looking and four backward-looking cameras\footnote{https://misr.jpl.nasa.gov}. The cameras are arranged so that the angles between
the lines of sight and the local vertical at the places of scanning are 26.1\degr, 45.6\degr, 60.0\degr, and 70.5\degr.
The swath width is only 380 km, which still encompasses the dust trail. There are four spectral bands.
The pixel resolution is 275 m for the red band of all nine cameras, as well as for all other bands of the nadir-viewing camera,
while the resolution of remaining bands (blue, green, and near-infrared) is 1.1 km for all eight off-nadir cameras. 
The scan time between the first forward-aimed camera and the last backward-aimed camera is about 6.5 minutes, 
therefore the individual cameras show the trail at different times.

All nine images of the dust trail and its shadow from MISR are displayed in Fig.~\ref{MISR}. The images are based
on a Level 1B2 product, where the data are provided projected on the standard ellipsoid. Because the dust trail was
located at much higher altitudes than the clouds, the parallax shift of the trail with respect to the clouds and the shadow 
depends on the viewing angle, thus differs significantly between the individual cameras. Nevertheless, 
the appearance of the trail changed not only because of viewing geometry and direction with respect to the Sun, 
but also because of physical changes in the trail itself that were due to atmospheric winds and trail diffusion.

The first image (upper left in Fig.~\ref{MISR}) was taken by the forward-looking camera with angle 70.5\degr, labeled Df.
The trail was scanned at about 23:51:34 UT from a distance of about 1650 km. The camera was looking nearly southward,
that is,\ at the Sun. The Sun-object-camera angle was 142\degr. In contrast to the nadir view from MODIS, the trail
is prolonged, but not perfectly straight. The lower part is dark because the non-illuminated side of the trail was visible.
This part of the trail was evidently optically thick. The upper part was optically thin and is bright due to scattered sunlight.

The middle image from the MISR nadir camera (labeled An) is more or less identical with the MODIS view. The last image
taken by the backward-looking 70.5\degr\ camera (Da) was scanned at about 23:58:08 UT, again from a distance of about 1650 km,
but looking north. The whole trail was illuminated. A faint part of the shadow extends toward the trail in this image.
The trail is wider than in the first image because of continuing diffusion 
and mixing with the air.

The last relatively fresh dust trail image is from the Advanced Baseline Imager (ABI) on board the GOES-17 satellite
operated by the National Oceanic and Atmospheric Administration (NOAA), USA. The geostationary satellite is located above
137.2\degr \ western longitude. In December 2018, the full disk was scanned every 15 minutes. The dust trail and its shadow
appeared in the 00:00 UT image (December 19). The remapped section of the image is shown in Fig.~\ref{GOES}.
This part was scanned at about 00:00:38 UT, that is,\ more than 12 minutes after the bolide.

\section{Bolide trajectory and orbit}
\label{trajectory}

The images of the dust trail and its shadow show that the trail initially was a long linear feature, albeit with different
dust concentration in different parts. Just after the bolide disappearance, the trail had to follow the bolide trajectory,
more exactly, the part of the trajectory where significant amount of dust was released. At later times, dust trails are
deformed and moved by high-altitude winds, their own buoyancy, and diffusion, as was documented in the case of Chelyabinsk
\citep[e.g.,][]{Chelya_Popova}. When we determine the initial position and orientation of the trail,
the bolide trajectory is reconstructed independently of the USGS data. Unfortunately, the
trail does not contain any information on the bolide speed. To compute the orbit, we had to rely on the USGS 
speed as the basis. 

\subsection{Location and orientation of the dust trail}

In principle, two images of fresh dust trail taken from two suitable widely separated sites would be
sufficient to triangulate the trail position. The problem is that the available images were taken with
some delay after the bolide, and the trail was deformed by the winds in the meantime. Moreover, the delay was different 
for different images. Direct triangulation was therefore not possible. We therefore used an approach
of modeling the trail motion starting from a variety of initial conditions. The modeled projections
of the trail and its shadow were then compared with the real images and the range of initial conditions
consistent with the observations was found.

\subsubsection{Method}

The initial conditions were defined by the longitude and latitude of the trail point at a height of 25 km above
the sea level, $\lambda_{25}, \varphi_{25}$, and the azimuth and zenith distance of the radiant, $A_R, z_R$.
The height 25 km was chosen because it is close to the height of bolide maximum reported by the USGS and the
trail surely existed at this height. Azimuth is counted from south to west here. The position of the
trail at another height $h$ (in km) and at a time $t$ (in s) elapsed after the bolide was computed simply as
\begin{eqnarray}
\lambda_h(t) &=& \lambda_{25}  + \Delta X /(111 \cos \varphi_{25}),     \\
\varphi_h(t) &=& \varphi_{25}  + \Delta Y /111,
\end{eqnarray}
where 111 km is the length of one degree in latitude, and the zonal and meridional differences in position are
\begin{eqnarray}
\Delta X &=& (25-h) \sin z_R \sin A_R  + U_h t ,    \\
\Delta Y &=& (25-h) \sin z_R \cos A_R  + V_h t.
\end{eqnarray}
Here $U_h$ is the zonal wind and $V_h$ is the meridional wind at the height $h$ in km s$^{-1}$.
Wind speeds at various heights were taken from the
European Centre for Medium-Range Weather Forecasts (ECMWF) model for 0 UT on December 19 (Table~\ref{wind}). 
We also checked the United Kingdom Met Office (UKMO) model for 12 UT of both days, and the values were similar.
The winds blew generally from southwestern or southern directions.

\begin{table}
\caption{ECMWF wind model at relevant heights, valid  for the bolide position and 0 UT on Dec 19, 2018}            
\label{wind}      
\centering                         
\begin{tabular}{c c c c}        
\hline\hline                
Pressure & Height &  Zonal wind ($U$) & Meridional wind ($V$)\\
hPa & km & m s$^{-1}$ & m s$^{-1}$ \\
\hline                    
1 & 48.4 & 39.0 & 16.4 \\ 
2 & 43.7 & 14.1 & 14.5  \\
3 & 39.9 &  $-4.6$ & ~~9.8  \\
5 & 35.6 & $-4.1$ & ~~9.3 \\ 
7 & 33.8 & ~~1.0 & 20.2 \\ 
10 & 31.3 & $-0.2$ & 17.3 \\ 
20 & 26.6 &~~7.7 & 13.9 \\ 
30 & 23.9  & ~~5.0 & 16.8 \\ 
50 & 20.7 & 15.7 & 13.9 \\ 
70 & 18.8 &~~9.2 & 11.2 \\ 
100 & 16.3& 15.0 & ~~9.8 \\ 
\hline                                   
\end{tabular}
\end{table}

The position of the shadow of the trail point at height $h$ was computed from
\begin{eqnarray}
\Delta X_s &=& \Delta X + (h-h_c) \sin A_\sun / \tan  \epsilon_\sun ,    \\
\Delta Y_s &=&\Delta Y + (h-h_c) \cos A_\sun / \tan  \epsilon_\sun,
\end{eqnarray}
where $h_c$ is the height of the cloud tops, $A_\sun$ is the azimuth of the Sun, and 
$\epsilon_\sun$ is the elevation of the Sun in the region of the shadow.

The computed trail positions were compared with the images from the MISR nadir camera, An, the two
outermost MISR cameras, Df and Da, and the Himawari and GOES images. 
We used the MINX (MISR INteractive eXplorer) software\footnote{https://github.com/nasa/MINX}, 
and added a coordinate grid to the MISR images.
The images were then rotated so that north was up. Similarly, a Mercator projection of the Himawari and 
GOES images was prepared, a coordinate grid was added, and the image was rotated. 
These images provided the dust trail as seen 
from various satellite positions projected on the surface of the World Geodetic System 1984 (WGS84) geoid. Table~\ref{satellites} provides
the coordinates of the satellites, elapsed times after the bolide, and the corresponding coordinates of the Sun
in the shadow region\footnote{Computed with the NASA JPL HORIZONS online service, https://ssd.jpl.nasa.gov/?horizons}. 
We neglected the fact that the individual images were not taken at one time instant but
were scanned for some time, and that the Terra satellite also moved during this time. The times and 
positions valid for scanning of the center of the trail were used.

\begin{table}
\caption{Times, satellite positions, and Sun coordinates for the analyzed trail images}            
\label{satellites}      
\centering                         
\begin{tabular}{c c c c c c c}        
\hline\hline                
Image & $t$ & $\lambda_{\rm sat}$ & $\varphi_{\rm sat}$ & $h_{\rm sat}$ & $A_\sun$ & $\epsilon_\sun$ \\
 & s & \degr & \degr & km & \degr & \degr  \\
\hline                    
An & 391 &  171.7 &57.09& 705&$-7.6$ &  8.30 \\
 Df & 194 & 180.9 & 68.40& 705& $ -8.4$ &8.25\\
Da & 588 &  166.4 &45.45& 705& $-6.9$  &8.35\\
Himawari & 135 &  140.7 &0.00 &35786& $-8.7$&8.20 \\
GOES-17 & 738 & 222.8 & 0.00 & 35786 & $-6.1$ & 8.40 \\
\hline                                   
\end{tabular}
\end{table}

The zonal and meridional distance of the projected point from the true point at height $h$ (i.e.,\ the parallax) is
\begin{eqnarray}
\delta X &=& h \sin A_{\rm sat} \tan z_{\rm sat} ,    \\
\delta Y &=& h \cos A_{\rm sat} \tan  z_{\rm sat},
\end{eqnarray}
where $A_{\rm sat}$ is the azimuth of the satellite and $z_{\rm sat}$ is the zenith distance of the satellite
as seen from the trail position $\lambda, \varphi$.  They are computed from
\begin{eqnarray}
\sin A_{\rm sat} &=&  \sin(\lambda-\lambda_{\rm sat}) \cos \varphi_{\rm sat} / \sin d ,   \\
\sin z_{\rm sat} &=&  (R+h_{\rm sat}) \sin d / r ,
\end{eqnarray}
where $\lambda_{\rm sat}$, $\varphi_{\rm sat}$, $h_{\rm sat}$ are coordinates of the satellite, 
$R$ is the radius of the Earth, $r$ is the distance of the satellite,
\begin{equation}
r=\sqrt{(R+h_{\rm sat})^2+(R+h)^2-2\ (R+h_{\rm sat})\ (R+h) \cos d},
\end{equation}
and $d$ is the angular distance of the trail and the satellite as seen from the Earth center,
\begin{equation}
\cos d=\sin \varphi \sin \varphi_{\rm sat} + \cos \varphi \cos \varphi_{\rm sat} \cos(\lambda-\lambda_{\rm sat}).
\end{equation}

The parallax correction was applied also to the cloud tops (i.e.,\ the shadow), which were approximately
at $h_c = 8.5$ km (as derived from the MODIS cloud-top height product).

   \begin{figure}
   \centering
   \includegraphics[width=\columnwidth]{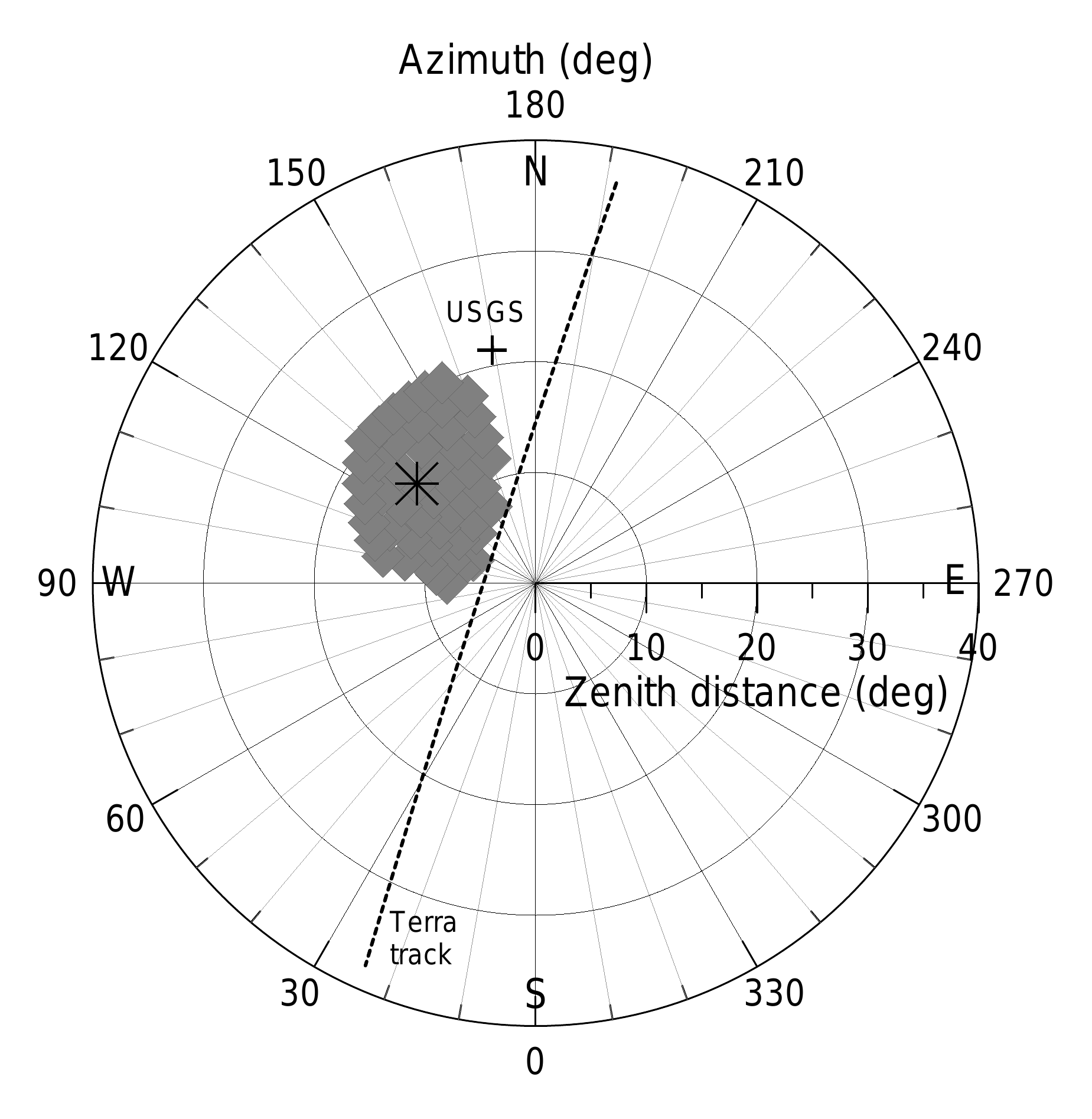}
   \caption{Polar plot of possible radiants of the bolide (gray area). The nominal radiant is given as an asterisk and the USGS
   radiant as a cross. The dashed line shows the apparent path of the Terra satellite as seen from the trail location.}
              \label{radiants}%
    \end{figure}

\subsubsection{Results}

The graphical comparison of the observed and computed trail projections yielded an unambiguous result concerning 
the location of the lower part (height = 25 km) of the trail:
\begin{eqnarray}
\lambda_{25} &=& 172.41\degr,  \nonumber  \\
\varphi_{25} &=& \ \ 56.88\degr.  \nonumber
\end{eqnarray}
 The uncertainty is $\sim$0.01\degr. This agrees well with the USGS data (172.4, 56.9), which were provided
 to only one decimal degree.
 
 The height span of the trail could also be determined well. The clearly visible dust trail extends from a height of
 23 km to 34 km. Less obvious traces of the trail and/or its shadow are visible in some images down to 18 km and up to 41 km.
 
    \begin{figure}
   \centering
   \includegraphics[width=0.70\columnwidth]{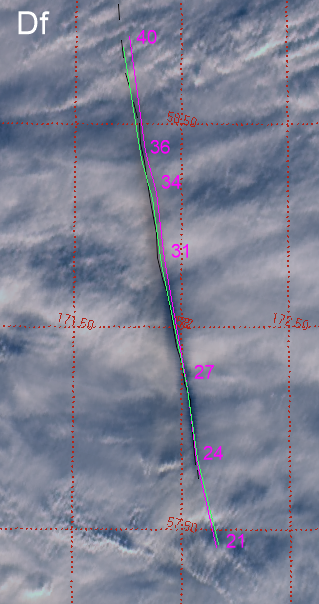}
   \caption{Image of the trail shadow from MISR camera Df, rotated and with the coordinate grid added. The center
   of the shadow was marked manually with a black line. The dashed line was used for indistinct parts of the shadow.
   The magenta line shows the modeled shadow assuming the USGS bolide radiant. The numbers are the corresponding
   trail heights (rounded heights for the pressure levels, where wind data are available; see Table~\ref{wind}).
   The green line shows the modeled shadow assuming the nominal radiant determined by us.}
              \label{proj1}%
    \end{figure}
 
The determination of the trail orientation, that is,\ the radiant of the bolide, was more problematic. The region 
of possible radiants on the sky has a diameter of about 15\degr\ (Fig.~\ref{radiants}).
All these radiants are consistent with the data. The boundaries of this region can be considered as the one-sigma limit. 
The USGS radiant lies just outside this limit. The most probable radiant (hereafter called the nominal radiant) has
an azimuth (counted from the south) and zenith distance:
\begin{eqnarray}
A &=& 130\degr,   \nonumber  \\
z &=& 14\degr. \nonumber
\end{eqnarray}
This radiant lies more to the west and closer to the zenith than the USGS radiant. In any case, the bolide
trajectory was steep, with a zenith angle somewhere in the range 7 -- 21\degr.

The bolide trajectory relative to the orbit of the Terra satellite was somewhat unfortunate.
The whole trajectory lay close to the orbital plane of the satellite. This is demonstrated by the fact that
the apparent path of the satellite as seen from the bolide location passed close to the zenith and that the bolide radiant 
was also close to the satellite path (see Fig.~\ref{radiants}). 
These circumstances, together with the fact that not the bolide itself
but only a diffuse and moving dust trail was observed, made the trajectory determination difficult.

Figure~\ref{proj1} shows the comparison of expected shadow position in the MISR Df camera image for the USGS and nominal radiant.
The USGS radiant produces the shadow of the upper part of the trail to be located slightly more to the east than observed, while the nominal
radiant agrees with the observation. From the shadow observations, we estimated the upper limit of radiant azimuth to be 160\degr,
which is possible for medium zenith distances (12--18\degr).

   \begin{figure}
   \centering
   \includegraphics[width=0.7\columnwidth]{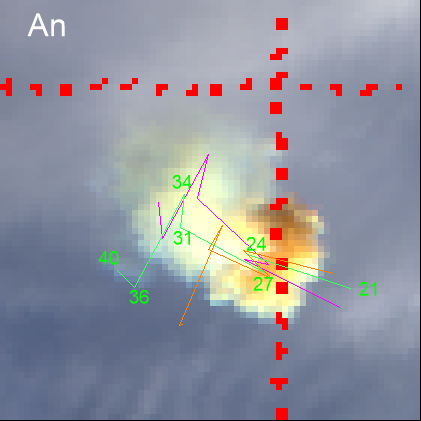}
   \caption{Enlarged and rotated image of the trail from MISR camera An. Red dots are part of the coordinate grid (meridian 172.5\degr,
   parallel 57\degr).  The magenta line shows the modeled trail projection assuming the USGS bolide radiant, the green line
   with height indications is for the nominal radiant, and the orange line is for vertical trajectory.}
              \label{proj2}%
    \end{figure}
    
   \begin{figure}
   \centering
   \includegraphics[width=0.48\columnwidth]{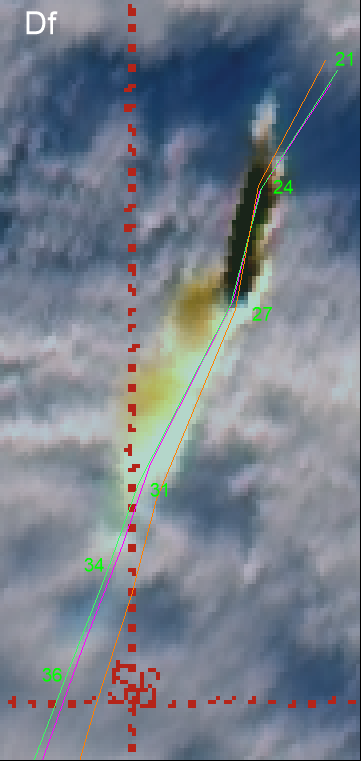}
      \includegraphics[width=0.48\columnwidth]{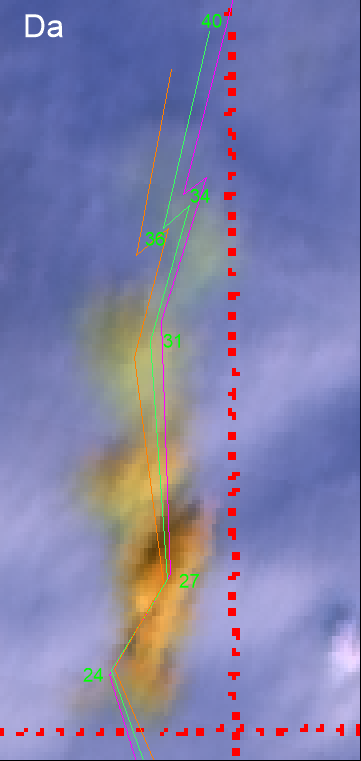}
   \caption{Enlarged and rotated image of the trail from MISR cameras Df and Da. Red dots are part of the coordinate grid (meridian 172\degr,
   parallel 56\degr\ on the left and 173\degr, 57.5\degr\ on the right). 
   The magenta line shows the modeled trail projection assuming the USGS bolide radiant, the green line
   with height indications is for the nominal radiant. The orange line is for vertical trajectory on the Df image 
   and for radiant with $A=90\degr, z=20\degr$ on the Da image.}
              \label{proj3a4}%
    \end{figure}

   \begin{figure}
   \centering
   \includegraphics[width=0.6\columnwidth]{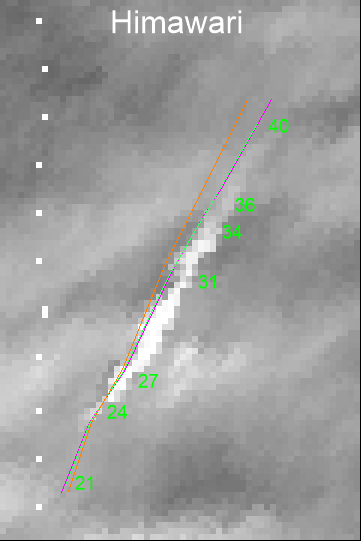}
   \caption{Enlarged and rotated image of the trail from the Himawari visual channel (0.64 $\mu$m), in Mercator projection. 
   White dots are part of the coordinate grid (meridian 173\degr).
   The magenta line shows the modeled trail projection assuming the USGS bolide radiant, the green line
   with height indications is for the nominal radiant, and the orange line is for radiant with $A=140\degr, z=26\degr$.}
              \label{proj5}%
    \end{figure}
    
       \begin{figure}
   \centering
   \includegraphics[width=0.7\columnwidth]{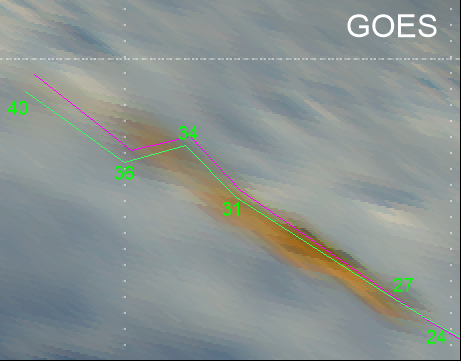}
   \caption{Enlarged and rotated image of the trail from GOES-17. 
   The magenta line shows the modeled trail projection assuming the USGS bolide radiant, the green line
   with height indications is for the nominal radiant.}
              \label{proj6}%
    \end{figure}

The vertical trajectory ($z=0\degr$) is not consistent with the data. It produces the shadow shifted to the east and, moreover, 
does not fit the trail images from cameras An and Df well. Figure~\ref{proj2} shows the enlarged view of the trail by camera An.
The computed projections for the USGS, nominal, and vertical radiants are shown as well. Because the trail is so wide, both the USGS and
the nominal solutions can be considered as consistent with the data. The dark part of the trail corresponds to
heights of about 24 -- 27 km (which were projected almost to the same place because of the existing wind pattern) and the bright part corresponds to heights above 30 km. The vertical trajectory, on the other hand, is not consistent with the data because the projected trail
is too short (it does not extend far enough westward). We estimated the possible minimal zenith distance of the radiant to be about 7\degr\
in case of radiant azimuths 90 -- 110\degr.

 Figure~\ref{proj3a4} (left) shows the enlarged view of the trail by the forward-looking camera Df. The USGS and nominal radiants give almost
 identical projections in this case. Here the optically thick part of the trail clearly spans heights $\sim$ 23.2 -- 26.6~km
 (see Table~\ref{wind} for more precise height levels than given in the figures).  The upper part is well visible up to a height of 31.5 km
 under this viewing conditions. The upper part seems to be not fitted well, but in fact we have no wind data between 26.6 -- 31.3~km. 
 In any case,  the prediction for the vertical trajectory is worse.
 
The western radiants ($A$ = 90 -- 100\degr) are possible only for small zenith distances ($z\leq15\degr$). For larger zenith distances, 
western radiants produce
 the shadow shifted to the west, the nadir projection is too long (going too far WSW), and the fit of the backward-looking camera Da 
is not very good at higher altitudes (see Fig.~\ref{proj3a4}, right). 
 
 The Da view shows the more evolved trail at a later time and under good illumination.
 Several dust concentrations lie at heights of about 24, 27, 29, and 31 km, as well as a fainter concentration at 34 km. Both the USGS and nominal
 radiants produce acceptable fits. 
  
 The shadow extends as far as the trail in the Da image (see Fig.~\ref{MISR}). 
 This part of the shadow is consistent with the shadow of the lower
 part of the trail, down to about 18 km. Its location is farther east than expected, meaning that the zonal wind was stronger 
 at 18.8 km than given in Table~\ref{wind}.
 This low part of the shadow is also visible in other MISR images, especially Ca and Ba, although less distinctly than in Da. Light-scattering
 conditions under different viewing angles are probably responsible for these differences. The trail itself is not visible below the height of 20~km
 in any image.

 Returning to the consideration of the bolide trajectory, the zenith angle could not be too large. The maximum value is $z\sim21\degr$
 for azimuths 130 -- 150\degr. The largest discrepancy for higher angles is obtained for the Himawari image. Figure~\ref{proj5} shows
 that $z = 26\degr$ is unacceptable because the upper part of the trail is projected farther west than observed.
 The USGS and nominal radiants produce almost identical projections from Himawari. They do not fit
 the observed trail perfectly, but at least go parallel with it. The reason for the small shift may be that the image in fact does not show
 the whole trail, but only its illuminated side.
 
 The GOES image was of little help for trajectory determination. Because the satellite was located in the direction of bolide flight, the
 computed trail orientation is not sensitive to the zenith angle. The variations in azimuth produce some variations, but for small zenith
 angles ($z<20\degr$), these variations are smaller than the width of the evolved trail, 12 minutes after the bolide.
 Figure~\ref{proj6} shows the comparison of the USGS and nominal trajectory projections. The nominal solution fits the trail somewhat better.

\subsection{Pre-atmospheric orbit}
 
The possible range of bolide radiants derived in the previous section was used to compute possible
heliocentric orbits of the bolide. The analytical method of \citet{Ceplecha} was used.
The bolide entry speed is also needed to compute the orbit. Because the trail images do not 
contain any information about the bolide speed, we had to rely on the speed reported by the USGS. 
As discussed in Sect.~\ref{USGS}, the reported speed 32.0 km s$^{-1}$ is unusually high, and such a high speed
was found to be in error in another event. The orbits were therefore also computed for a lower speed of 
27 km s$^{-1}$ to see how the speed uncertainty affects the orbit.

   \begin{table}
\caption{Geocentric radiant and heliocentric orbit (J2000.0) for the nominal radiant and speed  32 km s$^{-1}$.}            
\label{orbittable}  
\begin{tabular}{ll}        
\hline\hline        
Right ascension & 230\degr \\
Declination & 64\degr \\
Geocentric velocity         &  30 km s$^{-1}$ \\[0.5ex]
Semimajor axis & 1.8 AU \\
Eccentricity & 0.44 \\
Perihelion distance & 0.98 AU \\
Aphelion distance & 2.5 AU \\
Inclination & 53\degr \\
Argument of perihelion & 186\degr \\
Longitude of the ascending node & 266.75\degr \\
\hline                                   
\end{tabular}
\tablefoot{For the possible range of values, see Fig.~\ref{elements}.}
\end{table}

   \begin{figure*}
   \centering
   \includegraphics[width=1.5\columnwidth]{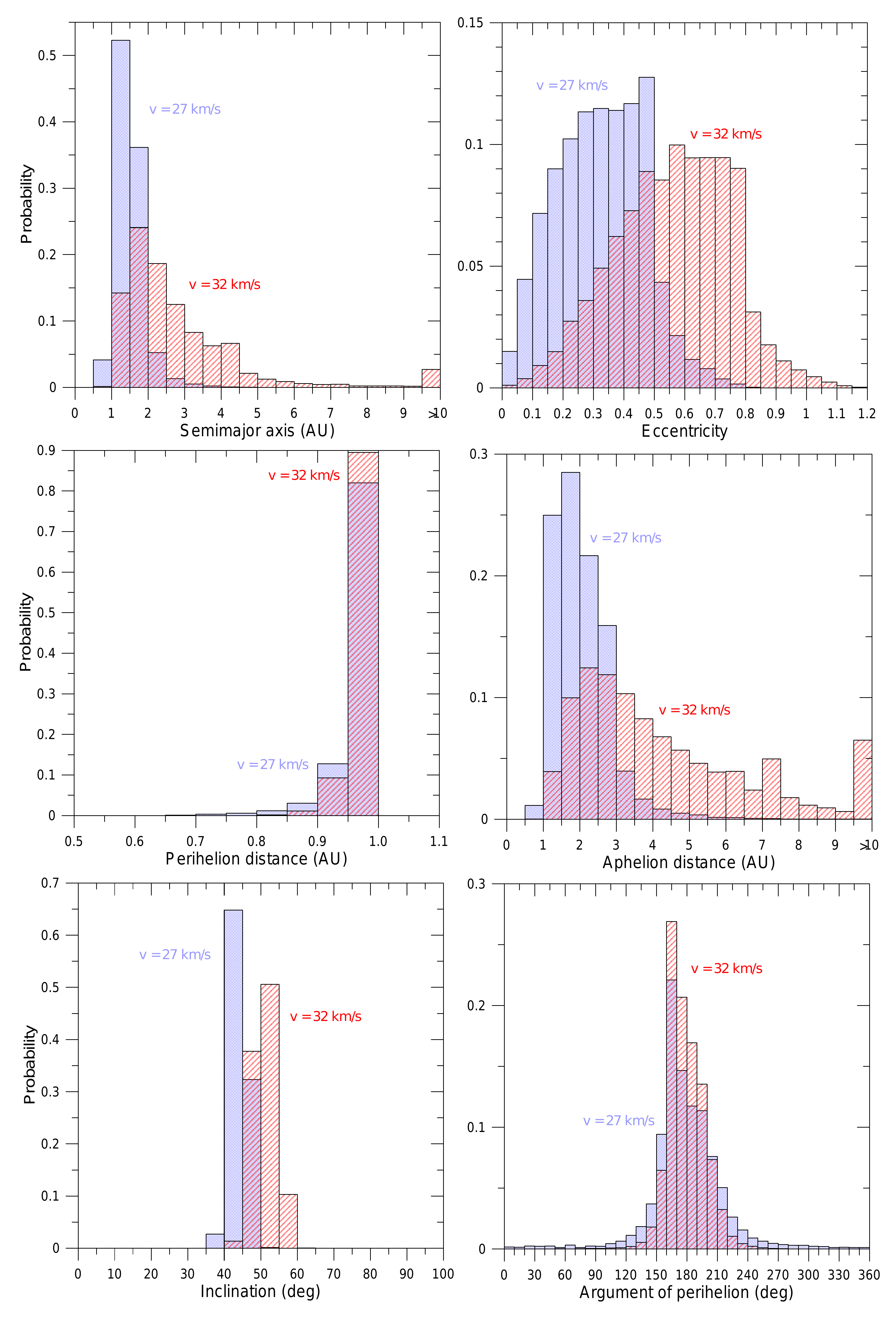}
   \caption{Probability distribution of orbital elements for two entry speeds, 32 km s$^{-1}$ (red) and 27  km s$^{-1}$ (blue). 
   The rightmost bars of the semimajor axis and aphelion distance plots sum probabilities for all values higher than 9.5 AU.}
              \label{elements}%
    \end{figure*}

The nominal orbit, computed for the nominal radiant determined from the dust trail and the USGS speed,
 is given in Table~\ref{orbittable}. The probability distribution of the values of
individual elements is presented graphically in Fig.~\ref{elements}. The computation was performed for a grid of
radiants with a step of 2\degr\ in azimuth and 1\degr\ in zenith distance. The boundaries of the shaded
area in Fig.~\ref{radiants} were taken as the one-sigma limit. All radiants up to 2.5 sigma were considered,
with the weight corresponding to the Gaussian distribution with the center in the nominal radiant. The computations
were made for two speeds.

Figure~\ref{elements} shows that some orbital elements are well constrained: the perihelion distance, inclination, and
the argument of perihelion. Their values lie in relatively narrow ranges for all possible radiants
and both speeds. There is a probability of about 90\% that the perihelion distance was between 0.95 -- 1 AU. 
More interestingly, the inclination was high in any case. With a probability of 90\%, it was between 45 -- 55\degr. 
Even when we admit the lower speed of 27 km s$^{-1}$, the inclination was almost certainly higher
than 40\degr. Finally, the argument of perihelion was close (within 30\degr) to 180\degr.
This result is connected with the fact that the encounter with the Earth occurred at the descending node, and
at the same time, close to the perihelion.

There are, on the other hand, wide ranges of possible values of the semimajor axis, eccentricity, and, consequently,
the aphelion distance. There is, nevertheless, an 68\% probability (for the speed 32 km s$^{-1}$) that the semimajor
axis was low, between 1 -- 3 AU. The eccentricity was most probably (72\%) between 0.4 -- 0.8. Asteroidal orbits with
aphelion distance $Q< 4.5$ AU have a probability of 64\%. Larger $Q$ are produced only by radiants at azimuths $A\gtrsim155\degr$
or zenith distances $z\lesssim5\degr$. Asteroidal orbits are therefore well possible for the speed of 32 km s$^{-1}$.
The situation changes dramatically with speed. For 27 km s$^{-1}$, virtually all orbits are asteroidal.

   \begin{figure*}
   \centering
   \includegraphics[width=18cm]{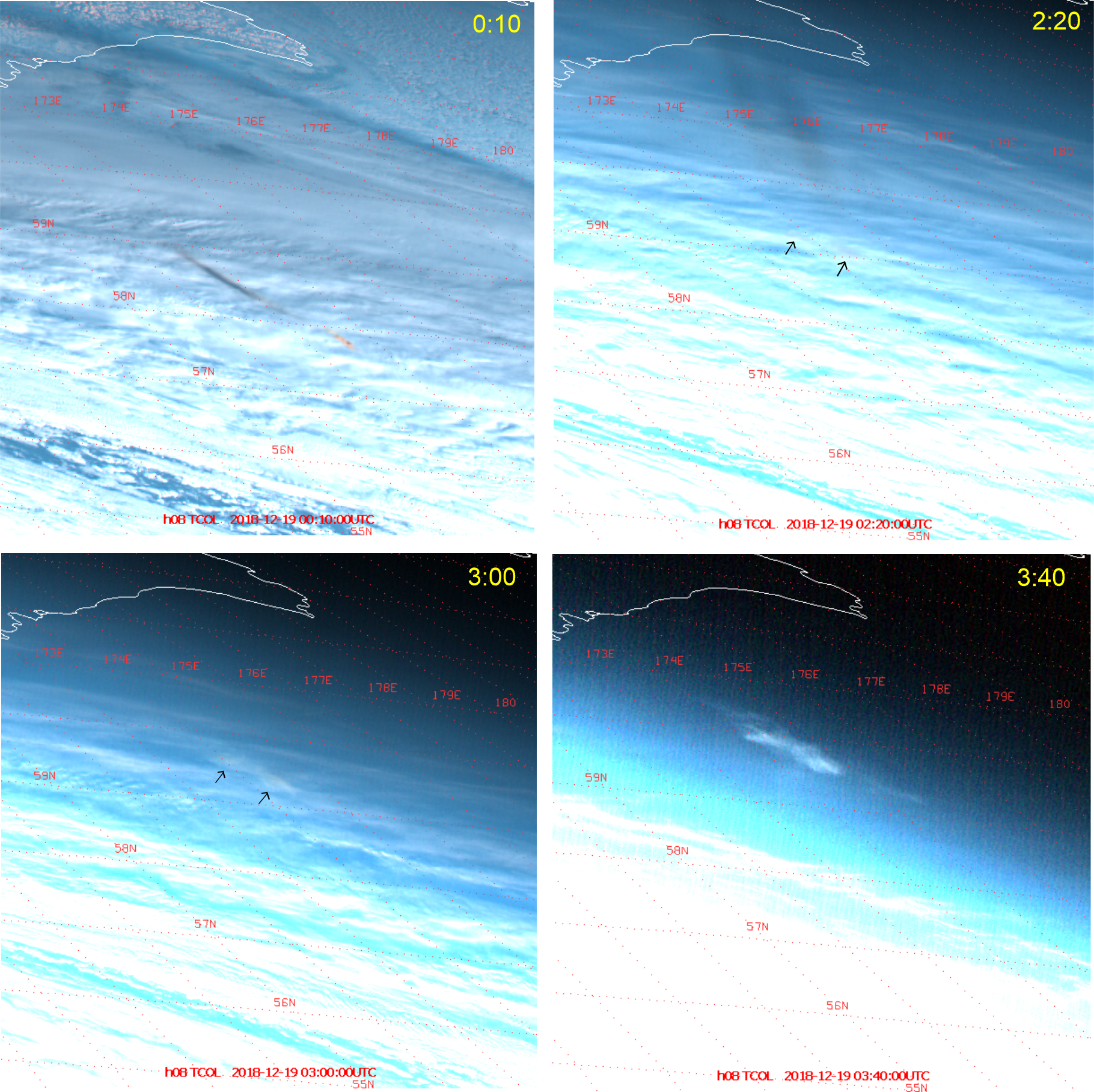}
   \caption{Four images of the dust trail obtained by the Himawari satellite between 0:10 and 3:40 UT on December 19. 
   The images are in RGB true color, and contrast is increasingly enhanced.}
              \label{Himawari-late}%
    \end{figure*}

\section{Trail at later times}
\label{evolution}

The trail and its shadow are traceable in the Himawari images until sunset, that is, in 25 images taken with 10-minute
intervals between 23:50 UT and 4:00 UT on December 19. Four of these images are shown in Fig.~\ref{Himawari-late}.
The dust continued its north-northeastward motion due to winds and gradually dispersed. The dust itself is hardly visible
in the 2:20 UT image, but three parallel long shadows face northeast. The Sun was low above
the southwestern horizon at that time. The shadow was no longer visible at 3:00 UT because the clouds farther northeast were
no longer illuminated at that time (i.e.,\ the shadow missed the Earth). 
The dust trail itself could already be better distinguished from the underlying clouds.
This was even better at 3:40 UT, when the clouds below the trail were no longer visible while the dust was still well illuminated. 
The dust was visible for the last time at 4:00 UT. The 4:10 UT image shows no signal at the trail position.

From the fact that the trail ceased to be visible after 4 UT, we roughly estimated the height of the two
brightest dust concentrations to be 25 -- 28 km. The lower part was located nearly at 173.8\degr\,E, 59.4\degr\,N,
that is,\ almost 300~km from the original bolide position. This position corresponds to an average speed of 20 m s$^{-1}$,
roughly consistent with the wind speed. No significant vertical motion of the dust was detected. The
highest dust concentration remained at the same height within the precision of the measurement for the whole
four hours.

   \begin{figure}
   \centering
   \includegraphics[width=0.9\columnwidth]{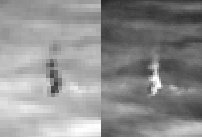}
   \caption{Enlarged monochromatic images of the dust trail obtained by the Himawari satellite at 23:50 UT. 
The blue Band 1 (0.45--0.49 $\mu$m) is at the left and the red Band 3 (0.60--0.68 $\mu$m) is at the right.
The spatial resolution of the red band is twice as high (0.5 km at the sub-satellite point) as in the blue band.}
              \label{Himawari-bands}%
    \end{figure}
    
       \begin{figure}
   \centering
   \includegraphics[width=\columnwidth]{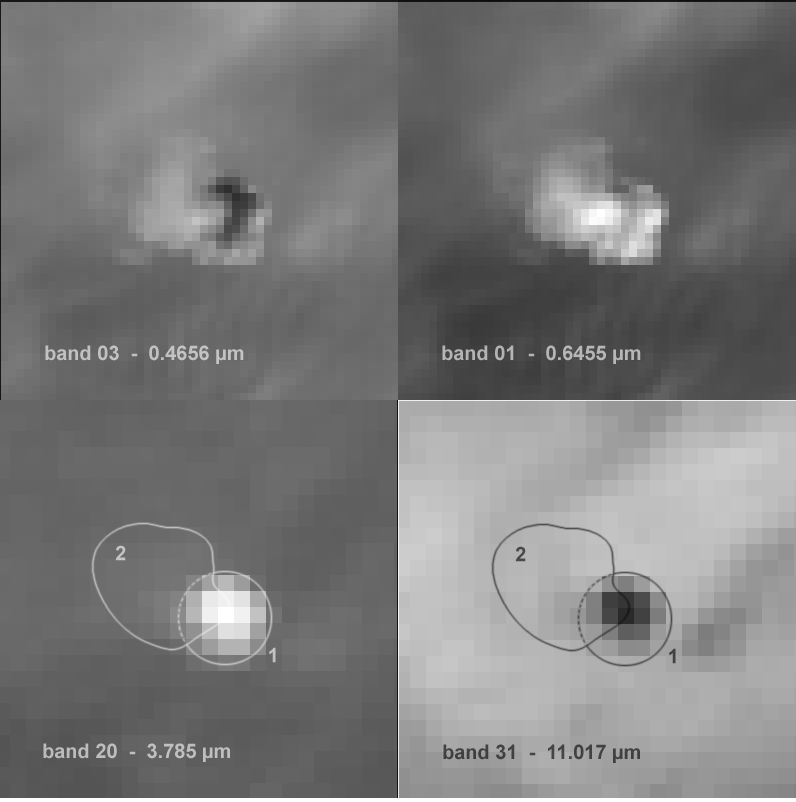}
   \caption{Enlarged monochromatic images of the dust trail obtained by the MODIS instrument on board the Terra satellite.
The blue Band 3 (0.46--0.48 $\mu$m) is at the upper left, the red Band 1 (0.62--0.67 $\mu$m) is at the upper right,
Band 20 (3.66--3.84 $\mu$m) is at lower left, and Band 31 (10.8--11.3 $\mu$m) is at lower right. The visible bands
have a resolution of 0.5~km, and the resolution of the infrared bands is 1~km. The circled area 1 is the dense lower part of the trail (heights 24--30 km),
area 2 is the upper part (30--34 km).}
              \label{MODIS-bands}%
    \end{figure}

\section{Spectral properties of the dust}
\label{spectrum}

In this section, we investigate the appearance of the dust trail and its shadow in various spectral bands. 
The most detailed information is available from the MODIS instrument on board the Terra satellite. MODIS has
36 spectral bands covering visible and infrared radiance from 0.4 $\mu$m to 14 $\mu$m. 
Generally, signal at short wavelengths is from reflected sunlight, while the long wavelength bands contain thermal emission.
Depending on the actual temperature of the object, some of the medium wavelengths bands contain both reflected
and emitted light.

The shadow was most pronounced at the shortest wavelengths, as expected.
This observation is in accordance with the fact that dust
absorption is inversely proportional to the wavelength. Only a short shadow, 
caused by the densest part of the trail, was
visible at 3.8 $\mu$m (Band 20).  
Of course, no shadow was visible at thermal wavelengths.

As we saw earlier, the dust trail itself had a reddish color (Fig.~\ref{Himawari}). There is a striking difference between
the blue and red bands in the Himawari imagery. The trail was dark in blue and bright in red (Fig.~\ref{Himawari-bands}).
MODIS provides a more detailed view from a different angle. Figure~\ref{MODIS-bands} shows a selection of four bands.
The shadowed part of the trail (brown in Fig.~\ref{detail}) was very dark in the blue band. This fact confirms that the
dust efficiently absorbed blue light. The directly illuminated parts of the trail were only marginally brighter in the
blue band than the underlying clouds.  Because blue light was not efficiently scattered by the dust, Rayleigh scattering did not 
occur, meaning that the majority of dust particles were comparable in size to or larger than the wavelength of the light. 

The red band image is not as dark in the shadow and it is much brighter in the illuminated parts. This means that
red light was more efficiently scattered or reflected and less absorbed than blue light.

The trail looks different in Band 20 (3.8 $\mu$m) image. The less dense upper section of the trail is invisible and 
the main section does not show any shadowed part. Instead, the trail is more or less symmetric, with the brightest pixel in the center.
There was little absorption at this wavelength. The reflection was most intensive in the center, most likely because 
of the highest particle concentrations in the trail core. The trail was much brighter
than the clouds, indicating much smaller particle size within the dust trail (most likely of the order of units of microns) than at the cloud tops (tens of microns). 

\begin{figure}
   \centering
   \includegraphics[width=\columnwidth]{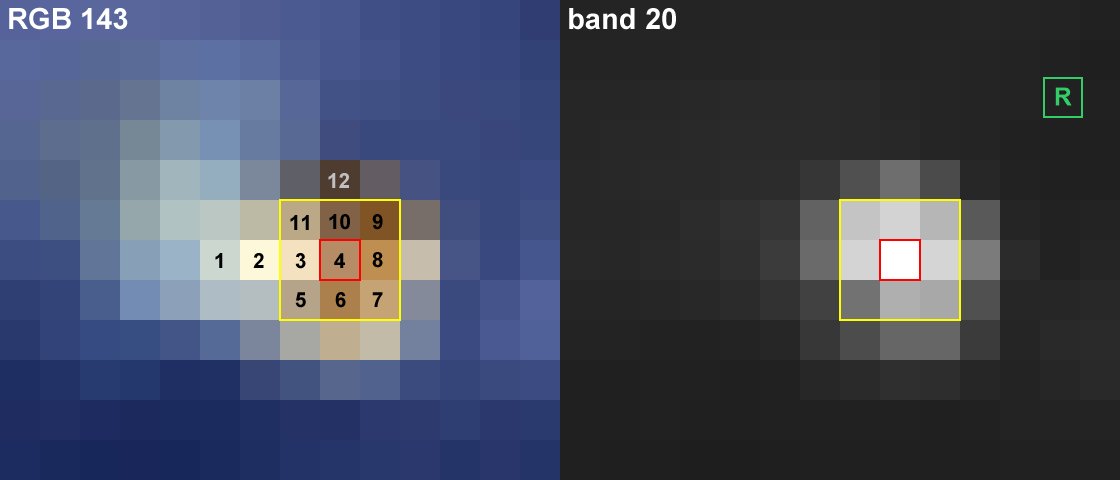}
   \caption{Visible light RGB composite (left) and Band 20 IR image of the trail in $1\times1$ km pixel resolution. Pixels
   with measured radiances are numbered. The reference pixel off the trail is marked R.}
              \label{pixnum}%
\end{figure}

In contrast, the trail was significantly darker than the clouds in Band 31 (11.0 $\mu$m). The trail therefore 
absorbed thermal radiation of the background clouds at this wavelength and did not emit too much by itself (due to its low temperature). 

\begin{figure}
   \centering
   \includegraphics[width=\columnwidth]{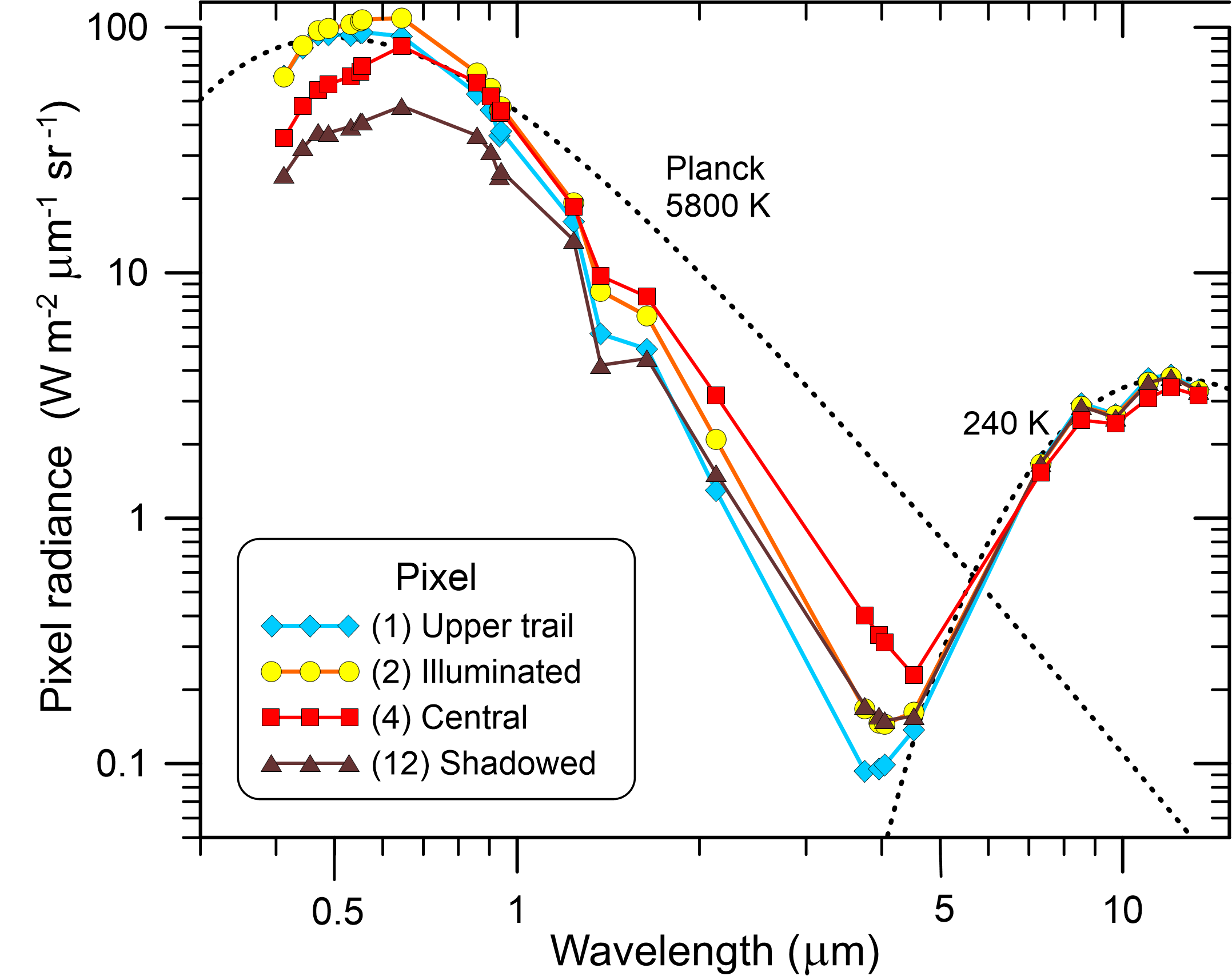}
   \caption{Radiances in MODIS bands as a function of wavelength in four selected representative pixels.
   See Fig.~\ref{pixnum} for pixel identification. Planck functions
   for 5800 K (solar color temperature) and 240 K (approximate cloud or dust temperature) are also shown.}
              \label{spectra}%
\end{figure}

To obtain more insight about the reflectance and absorption properties of the dust, we extracted radiances of selected pixels
from all MODIS bands. The pixels we used, converted into $1\times1$ km size in all bands, are marked in Fig.~\ref{pixnum}. 
Next, we used these radiances to plot low-resolution visible, near-infrared, and thermal infrared spectra. 
The spectra of four representative pixels are shown
in Fig.~\ref{spectra}. Bands 13--16, 21, 24, 27, and 34--36 were excluded because the signal was too noisy.
The plotted spectra belong to the upper thin trail (Pixel 1), the brightest pixel in the visible light (Pixel 2), the brightest
pixel in the reflected infrared light (Pixel 4, the center of the lower trail), and the darkest pixel in the visible light (Pixel 12).

\begin{figure}
   \centering
   \includegraphics[width=0.9\columnwidth]{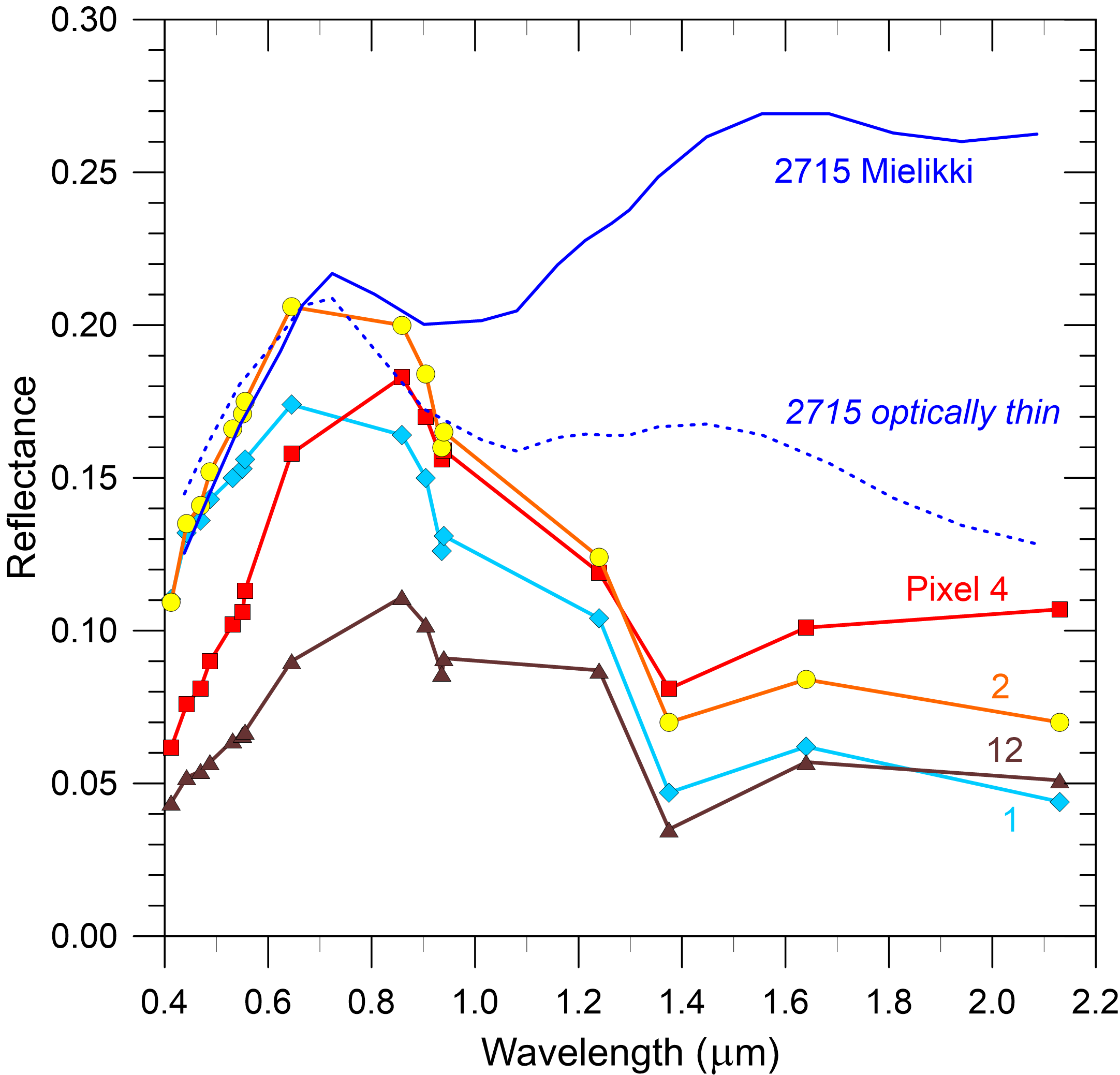}
   \caption{Reflectances in MODIS bands as a function of wavelength in four selected representative pixels. 
   See Fig.~\ref{pixnum} for pixel identification.
   The relative reflectance spectrum of asteroid 2715 Mielikki taken from \citet{DeMeo} is shown for
   comparison. The same Mielikki spectrum recomputed for the optically thin dust situation (see the text)
   is provided as a dotted line. The last two spectra were normalized at 0.65 $\mu$m to the same value as
   the pixel 2 spectrum.}
              \label{reflectance}%
\end{figure}

The spectra clearly show two components. Reflected solar spectrum dominates below 4 $\mu$m, including Band 20.
Thermal emission dominates at longer wavelengths. Thermal spectra are similar in all pixels and can be, to first approximation,
described by the Planck function at 240~K. The reflected spectrum does not correspond well to the solar color temperature
5800~K  and varies between pixels.
To study the spectrum of the reflected radiation better, radiances in the range 0.4 -- 2.2 $\mu$m were converted into
reflectances and are plotted in Fig.~\ref{reflectance}. 

An absorption by water vapor is visible at 1.4 $\mu$m. 
As the bolide trail was located in the dry stratosphere, we can instantly exclude absorption by environmental water vapor at these altitudes. 
This is no evidence for the presence
of water vapor in the dust trail. Because the dust was not optically thick at infrared wavelengths, each pixel also contains
some background radiation reflected from the underlying clouds. 
As these are mainly mid-level clouds, the upper-tropospheric water vapor present above them absorbs part of the incoming and 
reflected solar radiance, wherefore the clouds look darker in this band, and thus also the radiance within the dust plume 
appears lower than in the other bands.
This explanation is supported by the fact
that the absorption is strongest in pixel~1, which contains the smallest amount of dust. The water vapor absorption band in the spectrum
cannot be therefore used as an argument for cometary origin of the bolide. 
A smaller absorption by atmospheric water is also visible at 0.9~$\mu$m.

The effect of shadow cast by one part of the trail at its other parts is well visible by comparing the spectra of pixels~1 and~12.
While both pixels are similarly bright around 2 $\mu$m, the shadowed pixel~12 becomes progressively darker with decreasing 
wavelength. Partial shadowing also causes the central pixel~4 to look darker at short wavelengths than pixel~2.
Pixel~2 was fully illuminated and it is the best pixel to study the reflectance properties of the dust.

Pixel~2 reflectance in Fig.~\ref{reflectance} is conspicuous by the steep red slope between 0.4 -- 0.65 $\mu$m. 
The slope is less steep in pixels 1 and 12 because these pixels also contain some light from the underlying clouds.
The low reflectance in blue is the reason for the dark appearance of the trail in the blue band (Fig.~\ref{Himawari-bands}). 
 When compared with reflectance spectra of asteroids, similarly steep slopes are characteristic for asteroids
of spectral types A and L \citep{BusBinzel}. Figure~\ref{reflectance} also shows the spectrum of asteroid 2715 Mielikki, taken
from \citet{DeMeo}. This asteroid was classified as A-type by \citet{Burbine} and \citet{BusBinzel}, but \citet{DeMeo}
declined this classification on the basis of near-IR spectra without providing an alternative classification. Mielikki has the
same slope at 0.4 -- 0.65 $\mu$m as  pixel 2. Above 1 $\mu$m, however, the reflectance
of Mielikki is much higher than that of the dust trail. Typical A-type asteroids have even higher reflectance
in the infrared than Mielikki, and no asteroid type shows such decrease of reflectance in infrared as the dust trail.

The probable explanation is that the dust was optically thin in the infrared. When we consider a finite optical depth, $\tau_\lambda$,
the reflected spectrum can be expected to be $R_\lambda = r_\lambda(1-e^{-\tau_\lambda})$, where $r_\lambda$ is
the reflectance spectrum of the material (asteroid). Assuming, schematically, that optical depth is inversely proportional
to the wavelength, $\tau_\lambda = k/\lambda$, where $k$ is a constant, 
we can obtain a spectrum closer to the observation. Figure \ref{reflectance}
shows an optically thin Mielikki spectrum computed for $k=1\ \mu$m. Although the agreement with the observed spectrum is far from
perfect, the computed spectrum is much closer to the trail spectrum in the infrared than the asteroid spectrum. 
In the visible, the spectral slope, on the other hand, is somewhat shallower than observed. Finding a better agreement
would need detailed modeling of light scattering on the dust, which is beyond the scope of this work.

\begin{figure}
   \centering
   \includegraphics[width=0.85\columnwidth]{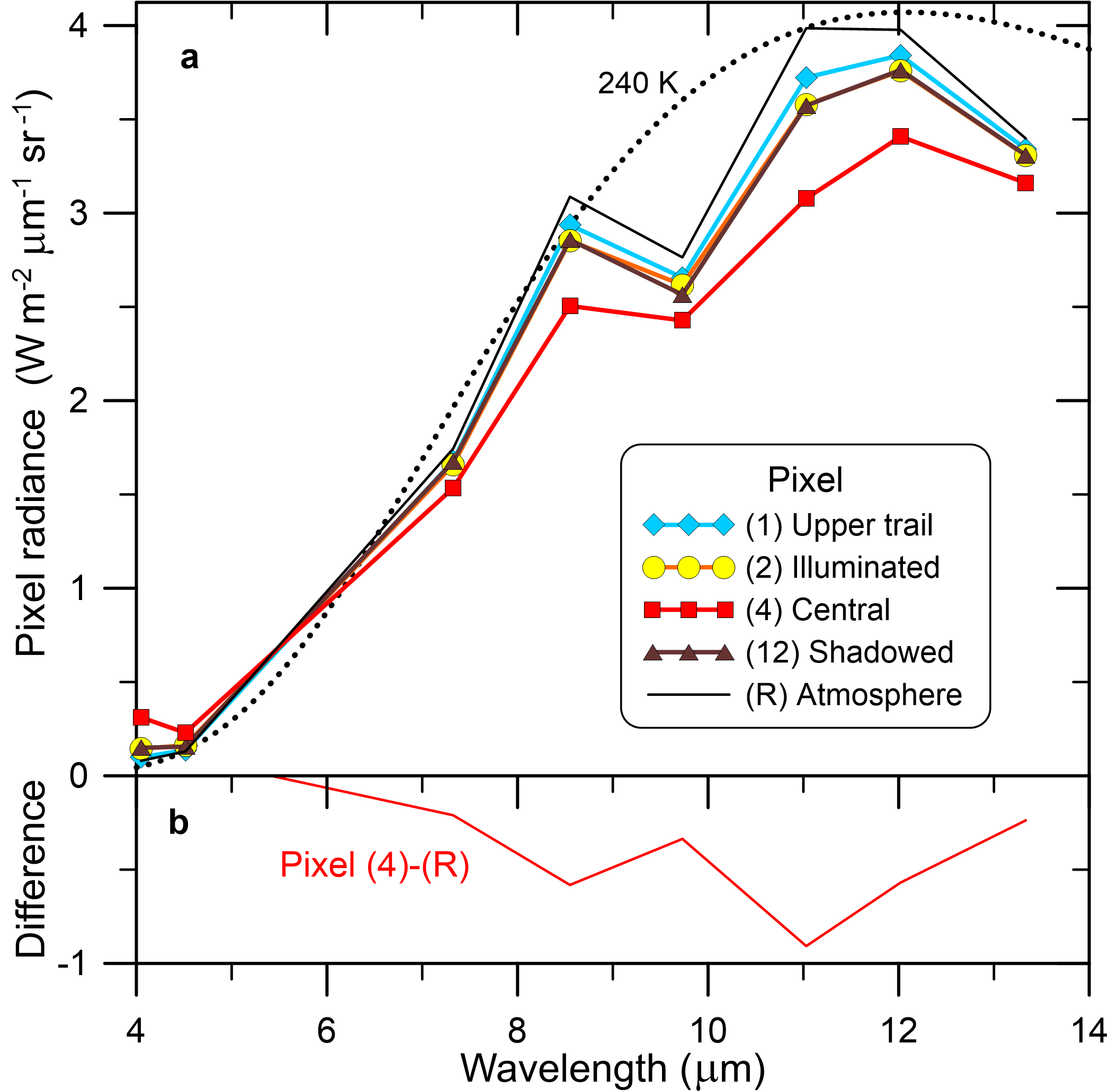}
   \caption{Radiances in the thermal part of the spectrum. The values for the reference pixel R containing only the atmosphere
   and the Planck function for 240 K are also shown. See Fig.~\ref{pixnum} for pixel identification. The lower panel (b)
   shows the difference of radiances in the central trail pixel 4 and the reference pixel R.}
              \label{IRspectrum}%
\end{figure}

The thermal infrared part of the trail spectrum is shown in Fig.~\ref{IRspectrum}. For comparison,  the spectrum of a reference
nearby pixel containing no dust is also given. The reference spectrum can be fit by a 240~K blackbody radiation, except for
the absorption by the stratospheric ozone at 9.7~$\mu$m (Band 30) and a lower radiance at 13.3~$\mu$m (Band 33), perhaps due to CO$_2$
absorption in the upper troposphere and stratosphere. All trail pixels were darker above 7~$\mu$m than the reference pixel. 
The trail therefore absorbed
cloud radiation. This effect has been observed for bolide dust before \citep{Charvat}. 
The absorption was proportional to the amount of dust and was most pronounced in pixel 4.

The absorption is mainly  due to the 10 $\mu$m silicate band. 
The silicate band at about 8 -- 12~$\mu$m contains two main maxima, one at 9.7~$\mu$m produced by
amorphous silicates, and one at 11.2~$\mu$m produced by crystalline olivine. Both maxima were observed in space, both in emission,
for example,\ in comet Hale Bopp \citep{HaleBopp}, and in absorption, for instance,\ in envelopes of young stellar objects \citep{YSO}.
In the bolide over Sudan \citep{Charvat}, fresh hot dust exhibited the amorphous peak in emission, while five minutes later,
the crystalline peak was observed in absorption. The production of the silicate band has traditionally been ascribed to submicron
dust particles, but recent work of \citet{Chornaya} showed that the band can also be produced by much larger particles
if they are irregular.

Figure~\ref{IRspectrum}b shows the absorption profile of pixel 4. 
The strongest absorption, about 25\%, occurred at 11~$\mu$m (Band 31). It is evidence
of the presence of crystalline olivine. The absorption was much weaker at 9.7~$\mu$m, but this band is affected by ozone absorption, so that
we cannot make any conclusions regarding amorphous olivine. It is possible that part of the atmospheric ozone was destroyed by the
bolide.

\section{Discussion and conclusions}
\label{discussion}

The analysis of space-borne images of the dust trail of the Bering Sea bolide enabled us to refine the atmospheric
trajectory of the bolide. Our radiant lies $13\degr \pm\ 9 \degr$ from that reported by the USGS, more to the west and closer to the zenith. 
The precision of radiant determination is lower than
for bolides that were directly recorded by cameras. The geographical location of the bolide is in agreement with the USGS data.
Nevertheless, we provide a precision of $\pm 0.01\degr$ instead of $\pm 0.1 \degr$. The reported height of the bolide
maximum brightness (25.6 km) is also in agreement with our data, as it lies
within the range of the highest dust concentration. We were unable to 
independently compute the bolide speed. The reported speed 32 km s$^{-1}$ is unusually high, but when combined with
our radiant, it provides reasonable asteroidal orbits, so that we consider it plausible.

The data analyzed by us did not allow us to compute the bolide energy. Nevertheless, \citet{Pilger} obtained 
50 kt TNT from infrasonic data, significantly less than the 173 kt TNT reported by the USGS. 
The bolide location obtained by them (57.86\degr\ N, 173.41\degr\ E) is different by more than
one degree from our location (56.88\degr\ N, 172.41\degr\ E) and cannot be correct. If their energy is correct,
the impacting asteroid was not so big as originally thought. With the speed of 32 km s$^{-1}$, the corresponding mass
is $4\times10^5$ kg. For an assumed density of 3500 kg m$^{-3}$, the diameter is 6 meters. 

The deep 
atmospheric penetration makes it plausible that the asteroid had a high density, corresponding to stony
meteorites. 
Meteoroids and asteroids that penetrate the atmosphere are subject to dynamic pressure on their leading side,
\begin{equation}
p= \rho v^2,
\end{equation}
where $\rho$ is the atmospheric density and $v$ is the speed. It was found that meteoroids repeatedly fragment under relatively
low pressures 0.1 -- 10 MPa, although the tensile strength of the material (stony meteorites) is about 30 MPa \citep{Popova, 2strengths}.
The low fragmentation strength is ascribed to cracks within the body.
The Chelyabinsk asteroid disrupted catastrophically under 1 -- 5 MPa and only small parts survived up to 18 MPa 
\citep{Chelya_Borovicka}.  Only the Carancas
crater-forming event was probably caused by a rare monolithic meteoroid with a size of about 1--2 m 
\citep{Carancas, Brown_Carancas}. 
Shower meteoroids in cometary orbits typically have low strengths of $<0.1$ MPa \citep{Trigo_strength, Shrbeny_EPS}.
The Bering Sea body probably fragmented several times, as evidenced by the
dust concentration at various heights. If the speed of 32 km s$^{-1}$ was kept at heights of 27 -- 25 km, as
the USGS data indicate, the dynamic pressure reached $\sim$ 30 MPa, however. This would suggest that a significant part of the asteroid 
(at least 10\% of the mass, i.e.,\ about half of the size, in order not to be decelerated much) was extraordinary strong, without cracks. 
Without more detailed data, it is difficult to make any conclusions about the asteroid fragmentation and strength, however.
Reaching the maximum brightness at the height of 25~km is not unusual;  a high speed above 30 km s$^{-1}$ at
this height is unusual.
In any case, the asteroid was not a fragile body of cometary origin.

       \begin{figure}
   \centering
   \includegraphics[width=\columnwidth]{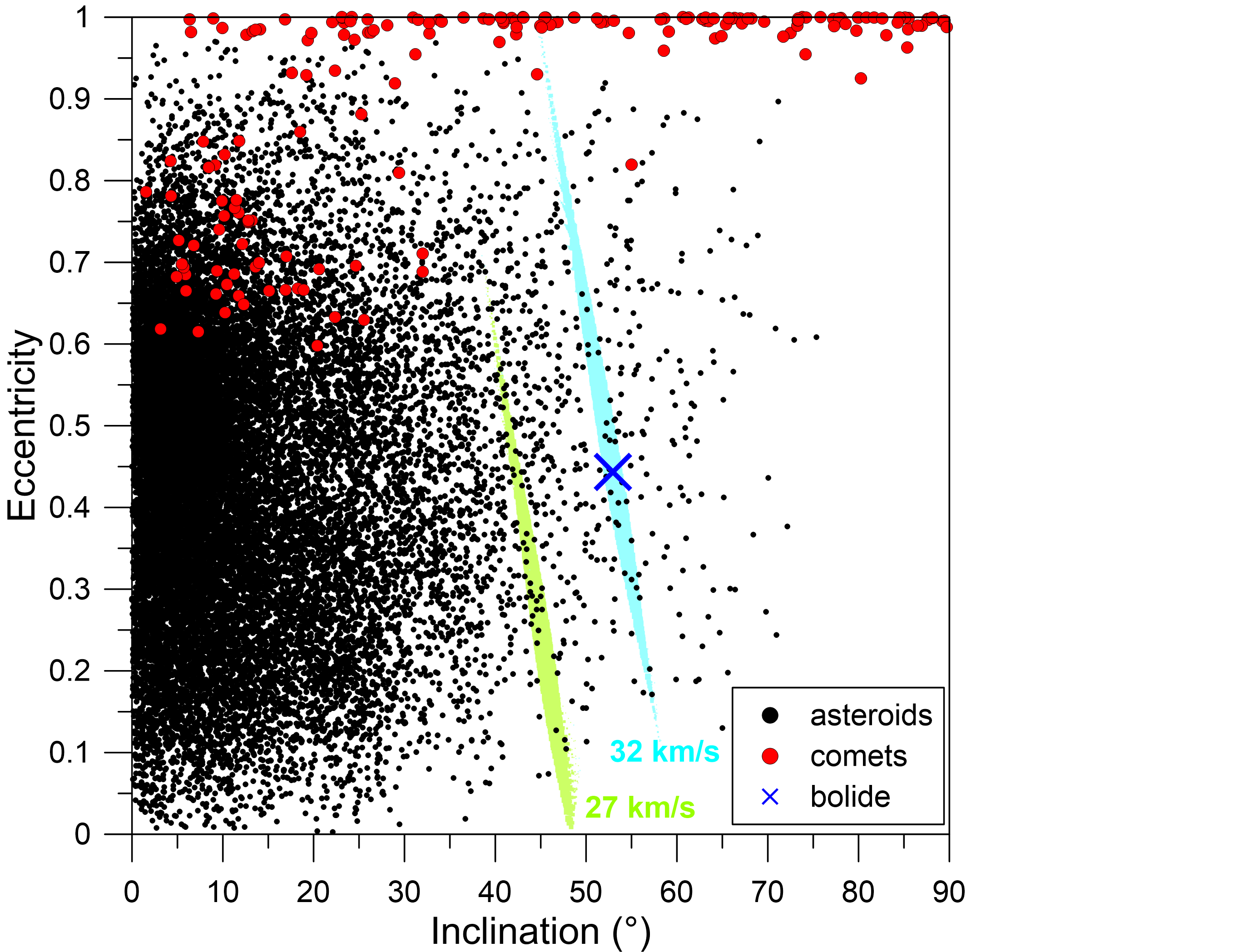}
   \caption{Eccentricity vs. inclination for inner Solar System ($q<1.2$ AU) asteroids and comets and the Bering Sea bolide.
   The nominal bolide solution is marked by the blue cross. The light blue area shows other possible radiant solutions within two sigma limits 
   for the speed of  32 km s$^{-1}$. The light green area is the same  for the assumed speed of 27 km s$^{-1}$.
   Asteroid and comet data are from the JPL Small-Body Database (https://ssd.jpl.nasa.gov/sbdb\_query.cgi).}
              \label{eiplot}%
    \end{figure}

The high inclination of the orbit could indicate a cometary origin. Nevertheless, the eccentricity-inclination plot (Fig.~\ref{eiplot})
shows that possible orbits, except for those with the highest eccentricities, fall in the area occupied by asteroids and not comets.
It is true that
the number of asteroids is much lower here than at lower inclinations. This can be partly a selection effect because asteroids
with high-inclination orbits have a lower probability of being discovered \citep{Tricarico}. Of the nine centimeter-sized iron meteoroids 
identified by \citet{Vojacek_iron}, three had high inclinations from 30\degr\ to 70\degr. The silicate absorption band
in the dust spectrum excludes the possibility that the Bering Sea asteroid was an iron object, however. The strong decrease in reflectance
from red to blue wavelengths suggests that the material was similar to that of asteroids of rare spectral types A or L. 
A-type asteroids are dominated by olivine and account for less than 0.16\% of all main-belt objects larger than 2 km \citep{DeMeo}.
L-type asteroids are probably even rarer and were found to be enriched by the material of calcium-aluminum-rich inclusions \citep{Devogele}.

Even if not as energetic as initially thought, the Bering Sea bolide was still one of the most energetic impacts in recent
decades. Without the monitoring service of the USGS, it would have remained unnoticed, although its signatures were present in the 
data of civilian Earth-observing satellites and infrasonic sensors. This fact demonstrates our limited ability to detect significant impact
events in remote areas of the globe. Until a dedicated global bolide detection system is constructed, it is advisable 
to scan other systems capable of detecting bolides for bolide signatures. 
The GLM bolide detection algorithm \citep{Rumpf} is a good example of such an initiative. The geographical coverage of lightning detectors
will be enlarged with the launch of the Meteosat Third Generation Lightning Imager \citep{LI}, 
which is currently scheduled for late 2022 or early 2023. We recall, however,
that such systems do not provide complete bolide data comparable to ground-based fireball networks, which, on the other hand,
have limited geographical coverage.

The Bering Sea asteroid was unusual because of its high entry speed, connected with the high orbital inclination. In terms of size, it
was not exceptional. A six-meter body is expected to impact Earth every second year on average. On the other hand, it was large enough 
to be accessible telescopically in space in the hours before impact. This event was favorable in the sense of solar elongation (about
90\degr) and because the body never entered the Earth shadow. 
However, it was moving at high northern declinations, where asteroid surveys are usually not performed. In principle, 
a casual detection by other systems was also possible.
During 1 -- 2 hours before impact, the object was located about 30\degr\ above NNE horizon as seen from Europe, on the night sky.
The apparent magnitude could be 13 -- 15 at that time. No detection was reported so far.
The lesson for planetary defense is that the threat can come from an unexpected part of the sky.

\begin{acknowledgements}
We thank Peter Brown for drawing our attention to the event and for providing us the ECMWF wind data.
This work used data from NASA Terra satellite (MODIS and MISR instruments), Japanese Space Agency Himawari satellite
(AHI instrument), and NOAA GOES-17 satellite (ABI instrument).
The MISR data were obtained from the NASA Langley Research Center Atmospheric Science Data Center.
We also acknowledge the use of the JPL HORIZONS System.
JB was supported by grant no.\ 19-26232X from Czech Science Foundation.
The work of MS was carried out under support of the Czech Ministry of Environment, DKRVO, CHMI 2018-2022 program.
\end{acknowledgements}

\end{document}